\documentclass[11pt]{article}
\usepackage{jheppub}

\usepackage{amsmath}
\usepackage{amssymb}
\usepackage{graphicx}

\newcommand{\bea}{\begin{eqnarray}}
\newcommand{\eea}{\end{eqnarray}}
\newcommand{\ft}[2]{{\textstyle\frac{#1}{#2}}}
\newcommand{\fft}[2]{\frac{#1}{#2}}
\newcommand{\nn}{{\nonumber}}

\preprint{MCTP-13-15}

\title{\boldmath Higher derivative corrections to Lifshitz backgrounds}

\author{Gino Knodel}
\author{and James T. Liu}

\affiliation{Michigan Center for Theoretical Physics, Randall Laboratory of Physics,\\
The University of Michigan, Ann Arbor, MI 48109--1040, USA}

\emailAdd{gknodel@umich.edu}
\emailAdd{jimliu@umich.edu}

\abstract{We explore the effect of curvature-square corrections on Lifshitz solutions
to the Einstein-Maxwell-dilaton system.  After exhibiting the renormalized
Lifshitz scaling solution to the system with parameterized $R^2$ corrections,
we turn to a toy model with coupling $g(\phi)C_{\mu\nu\rho\sigma}^2$ and
demonstrate that such a term can both stabilize the dilaton and resolve the
Lifshitz horizon to AdS$_2\times\mathbb R^2$.  As an example, we construct
numerical flows from AdS$_4$ in the UV to an intermediate Lifshitz region
and then to AdS$_2\times\mathbb R^2$ in the deep IR.}

\begin{document}
\maketitle
\flushbottom


\section{Introduction}

The application of holographic techniques to condensed matter systems
has led to the study of non-relativistic fixed points invariant under
the non-relativistic scaling 
\begin{equation}
t\to\lambda^{z}t,\qquad\vec{x}\to\lambda\vec{x},\label{eq:Lifshitz scaling symmetry}
\end{equation}
where $z$ is the dynamical exponent. Holographically, this scaling
may be realized by taking a bulk metric of the form \cite{Kachru:2008yh}%
\footnote{There are, of course, several equivalent ways of writing this metric, and
we will make use of this freedom below when investigating the Lifshitz flows in
section~\ref{sec:smoothing}.}
\begin{equation}
ds_{d+1}^{2}=-e^{2zr/L}dt^{2}+e^{2r/L}d\vec{x}^{2}+dr^{2},
\label{eq:lifbulk}
\end{equation}
where $r$ is a radial coordinate, and $L$ sets the length dimension
in the bulk. The scaling (\ref{eq:Lifshitz scaling symmetry}) is then accompanied by
the transformation
\begin{equation}
r\to r-L\log\lambda.
\label{eq:Lss2}
\end{equation}
This background is generally referred to as a Lifshitz
spacetime, and it has been the subject of much recent interest.

Exact Lifshitz geometries were constructed in \cite{Taylor:2008tg}
based on a simple model of a massive vector field coupled to Einstein
gravity. Turning on the time component of the vector breaks $d$-dimensional
Lorentz symmetry and gives rise to a family of backgrounds with $z\ge1$.
Alternatively, Lifshitz backgrounds may be obtained in the near horizon
region of dilatonic branes. A simple realization is to take an Einstein-Maxwell-dilaton
system of the form
\cite{Goldstein:2009cv,Cadoni:2009xm,Chen:2010kn,Charmousis:2010zz,Perlmutter:2010qu,Goldstein:2010aw}
\begin{equation}
e^{-1}\mathcal{L}=R-\ft12(\partial\phi)^{2}-f(\phi)F_{\mu\nu}F^{\mu\nu}-V(\phi).\label{eq:emdlag}
\end{equation}
Lifshitz scaling is obtained by taking a single exponential for the
gauge kinetic function along with a constant potential%
\footnote{Backgrounds dual to systems exhibiting hyperscaling violation may
be obtained by instead taking an exponential potential.%
} 
\begin{equation}
f(\phi)=e^{\lambda_{1}\phi},\qquad V(\phi)=-\Lambda.
\end{equation}
 The scaling solution has a running dilaton and a dynamical exponent
given by the relation 
\begin{equation}
\lambda_{1}^{2}=\fft{2(d-1)}{z-1}.
\end{equation}
 In addition, full solutions may be constructed that interpolate between
AdS$_{d+1}$ in the UV and Lifshitz in the IR.

As a consequence of the running dilaton, the Lifshitz solution runs
into strong coupling either in the UV for the electrically charged
solution or the IR for the magnetic solution (in the case $d=3$).
For the magnetic case, the possibility of quantum corrections was
investigated in \cite{Harrison:2012vy} by constructing a toy model
where the gauge kinetic function picks up an expansion in the effective
coupling $g\equiv e^{-\frac{1}{2}\lambda_{1}\phi}$ 
\begin{equation}
f(\phi)=\fft1{g^{2}}+\xi_{1}+\xi_{2}g^{2}+\cdots.
\end{equation}
 Under appropriate conditions, these loop corrections will stabilize
the dilaton and lead to the emergence of an AdS$_{2}\times\mathbb{R}^{2}$
geometry in the deep IR. The emergence of this AdS$_{2}\times\mathbb{R}^{2}$
region also has the benefit of resolving the Lifshitz horizon, which
would otherwise lead to tidal singularities
\cite{Kachru:2008yh,Copsey:2010ya,Horowitz:2011gh}%
\footnote{Note that the dilaton can also be stabilized in the dyonic case
\cite{Goldstein:2009cv,Goldstein:2010aw}, as well as in models with multiple Maxwell
fields \cite{Tarrio:2011de,Mozaffar:2012bp}.}.

In contrast with the magnetic solution, the electric solution ought
not to pick up quantum corrections in the IR, as the dilaton runs
to weak coupling. In this case, the Lifshitz horizon would not get
resolved by the same mechanism. However, in a stringy context (or
in that of any UV complete theory of gravity), there is another potential
source of corrections that arise from higher curvature terms. Although
Riemann invariants remain finite at the tidal singularity, this singularity
is nevertheless felt by strings \cite{Horowitz:2011gh}. Hence the
Lifshitz horizon would presumably be resolved in a consistent manner
in a stringy realization.

In this paper, we investigate the possibility that higher curvature
terms can resolve the Lifshitz horizon into an AdS$_{2}$ region in
the deep IR. In particular, we add $R^{2}$ terms to the Einstein-Maxwell-dilaton
system (\ref{eq:emdlag}) and seek electrically charged brane solutions
that flow from AdS$_{d+1}$ in the UV to Lifshitz and then to AdS$_{2}\times\mathbb{R}^{d-1}$
in the deep IR. As demonstrated in \cite{Adams:2008zk}, higher curvature
terms do not necessarily destroy the Lifshitz scaling solution, but
simply renormalize the dynamical exponent $z$. Thus we expect that
brane solutions with a large intermediate Lifshitz region do exist.
However, whether such solutions will flow smoothly into AdS$_{2}\times\mathbb{R}^{d-1}$
will depend on the parameters of the model. We investigate the $d=3$
case in some detail below, and in particular we confirm numerically
that smooth flows do exist that interpolate from AdS$_{4}$ to Lifshitz
to AdS$_{2}\times\mathbb{R}^{2}$.

This paper is organized as follows. In section~\ref{sec:hdlif},
we extend the Einstein-Maxwell-dilaton system by adding parameterized
$R^{2}$ corrections and construct the resulting renormalized Lifshitz
solutions. Since these solutions involve a running dilaton, we demonstrate
in section~\ref{sec:smoothing} that the dilaton can be stabilized
by introducing a dilatonic coupling to $R^{2}$. The resulting geometry
then takes the form AdS$_{2}\times\mathbb{R}^{d-1}$ in the deep IR.
Finally, in section~\ref{sec:discussion}, we conclude with a discussion
on some open issues.


\section{Lifshitz solutions in higher derivative gravity}
\label{sec:hdlif}

Lifshitz solutions in the presence of higher curvature terms were previously
investigated in
\cite{AyonBeato:2009nh,Cai:2009ac,Pang:2009pd,AyonBeato:2010tm,Dehghani:2010kd,Dehghani:2010gn,Brenna:2011gp,Lu:2012xu}.
Here, we focus on the Einstein-Maxwell-dilaton system, (\ref{eq:emdlag}), and
take the potential to be a constant, $V(\phi)=-\Lambda$, so that Lifshitz scaling
may be obtained at the two-derivative level.
The first set of corrections occurs at the four-derivative level, and in the gravitational sector
may be parameterized by three constants, $\alpha_1$, $\alpha_2$ and $\alpha_3$, where
the action is given by
\begin{equation}
S=\int d^{d+1}x\sqrt{-g}\left(R+\Lambda-\tfrac{1}{2}\left(\partial\phi\right)^{2}-f(\phi)F_{\mu\nu}F^{\mu\nu}+\alpha_{1}R_{\mu\nu\rho\sigma}R^{\mu\nu\rho\sigma}+\alpha_{2}R_{\mu\nu}R^{\mu\nu}+\alpha_{3}R^{2}\right).\label{eq:general action}
\end{equation}
Using Bianchi identities, we may write Einstein's equations as:
\begin{eqnarray}
T_{\mu\nu} & = & R_{\mu\nu}-\frac{1}{2}g_{\mu\nu}R+2\alpha_{1}R_{\mu\rho\lambda\sigma}R_{\nu}^{\phantom{\nu}\rho\lambda\sigma}+(4\alpha_{1}+2\alpha_{2})R_{\mu\rho\nu\lambda}R^{\rho\lambda}-4\text{\ensuremath{\alpha}}_{1}R_{\mu\rho}R_{\nu}^{\rho}\nn\\
 &  & -(2\alpha_{1}+\alpha_{2}+2\alpha_{3})\nabla_{\mu}\nabla_{\nu}R+(4\alpha_{1}+\alpha_{2})\Box R_{\mu\nu}+2\alpha_{3}RR_{\mu\nu}\nn\\
 &  & -\frac{1}{2}g_{\mu\nu}\left[\alpha_{1}R_{\rho\lambda\sigma\kappa}R^{\rho\lambda\sigma\kappa}+\alpha_{2}R_{\mu\nu}R^{\mu\nu}+\alpha_{3}R^{2}-(\alpha_{2}+4\alpha_{3})\Box R\right],
 \label{eq:general einstein equation}
\end{eqnarray}
where the energy momentum tensor is given by
\begin{equation}
T_{\mu\nu}=\frac{1}{2}\partial_{\mu}\phi\partial_{\nu}\phi+2f\left(\phi\right)(F_{\mu}^{\phantom{\mu}\rho}F_{\nu\rho}-\frac{1}{4}g_{\mu\nu}F_{\rho\sigma}F^{\rho\sigma})+\frac{1}{2}g_{\mu\nu}(\Lambda-\frac{1}{2}\partial^{\rho}\phi\partial_{\rho}\phi).\label{eq:general em tensor}
\end{equation}
These equations need to be supplemented with the equations of motion
of $F_{\mu\nu}$ and $\phi$: 
\begin{eqnarray}
\nabla_\mu(f(\phi)F^{\mu\nu})& = & 0\label{eq: F eom},\\
\square\phi-f^{\prime}(\phi)F_{\mu\nu}F^{\mu\nu} & = & 0,\label{eq:dilaton eom}
\end{eqnarray}
where $f'(\phi)$ is the derivative of $f(\phi)$ with respect to $\phi$.

Our goal is to find a matter field background that supports the Lifshitz metric,
(\ref{eq:lifbulk}).  From here on, we set $L=1$ without loss of generality.  Thus we
have
\begin{equation}
ds_{d+1}^2=-e^{2zr}dt^2+e^{2r}d\vec x^2+dr^2,
\label{eq:lifmet}
\end{equation}
and we want to determine the form of $F_{\mu\nu}$ and $\phi$.  We first note that
Maxwell's equations, \eqref{eq: F eom}, can be integrated to obtain an electric solution:
\begin{equation}
F=\frac{Q}{f(\phi)}e^{(z-(d-1))r}dr\wedge dt,\label{eq:F Ansatz}
\end{equation}
where $Q$ is an integration constant (the electric charge). Note that we
allow $\phi$ to depend on $r$ only. The components of the energy
momentum tensor are then given by
\begin{eqnarray}
T_{00} & = & g_{00}\left(-\frac{Q^{2}}{f(\phi)}e^{-2(d-1)r}-\frac{1}{4}(\phi^{\prime})^{2}+\frac{\Lambda}{2}\right),\nonumber \\
T_{rr} & = & g_{rr}\left(-\frac{Q^{2}}{f(\phi)}e^{-2(d-1)r}+\frac{1}{4}(\phi^{\prime})^{2}+\frac{\Lambda}{2}\right),\nonumber \\
T_{ij} & = & g_{ij}\left(\frac{Q^{2}}{f(\phi)}e^{-2(d-1)r}-\frac{1}{4}(\phi^{\prime})^{2}+\frac{\Lambda}{2}\right).\label{eq:EM Tensor for electric case}
\end{eqnarray}
Invariance of $T_{\mu\nu}$ under Lifshitz scaling requires $\phi\propto r$
and $f^{-1}\propto e^{2(d-1)r}$. More explicitly, we may rewrite
Einstein's equations, (\ref{eq:general einstein equation}), as:
\begin{eqnarray}
(\phi^{\prime})^{2} & = & 2(e^{-2zr}\mathrm{RHS_{00}}+\mathrm{RHS_{rr}}),\label{eq:phi' in terms of RHS}\\
\Lambda & = & \mathrm{RHS_{rr}}+e^{-2r}\mathrm{RHS_{ii}},\label{eq:Lambda in terms of RHS}\\
\frac{Q^{2}}{f(\phi)}e^{-2(d-1)r} & = & \frac{1}{2}(e^{-2zr}\mathrm{RHS_{00}}+\mathrm{RHS_{ii}}).\label{eq:Q^2 in terms of RHS}
\end{eqnarray}
The right hand side of each equation is a fourth order polynomial in $z$ and does
not depend on $r$. (The curvatures are computed in Appendix~\ref{app:A}.)
After integrating out the electric field, the dilaton equation of motion reads
\begin{equation}
\phi^{\prime\prime}(r)+(d-1)\phi'(r)+2\frac{f^{\prime}(\phi)}{f(\phi)}\frac{Q^{2}}{f(\phi)}e^{-2(d-1)r}=0.\label{eq:dilaton eom for electric case}
\end{equation}
Plugging in $f\propto e^{-2(d-1)r}$ and recalling that $\phi$ is
linear in $r$, we now find that the gauge kinetic function has to be
a single exponential $f\left(\phi\right)=e^{\lambda_{1}\phi}$.

Before we write down the final solution, let us change to a more convenient
basis of higher derivative terms by writing the corresponding Lagrangian as 
\begin{equation}
\mathcal{L}_{\mathrm{hd}}=\alpha_{{\scriptscriptstyle \mathrm{W}}}C_{\mu\nu\rho\sigma}C^{\mu\nu\rho\sigma}+\alpha_{\mathrm{{\scriptscriptstyle GB}}}G+\alpha_{{\scriptscriptstyle \mathrm{R}}}R^{2},\label{eq:hd-Lagrangian, new basis}
\end{equation}
with the Weyl tensor 
\begin{equation}
C_{\mu\nu\rho\sigma}=R_{\mu\nu\rho\sigma}-\frac{1}{d-1}\left(g_{\mu[\rho}R_{\sigma]\nu}-g_{\nu[\rho}R_{\sigma]\mu}\right)+\frac{1}{d\left(d-1\right)}g_{\mu[\rho}g_{\sigma]\nu}R,\label{eq:Weyl-tensor definition}
\end{equation}
and the Gauss-Bonnet combination
\begin{equation}
G=R_{\mu\nu\rho\sigma}R^{\mu\nu\rho\sigma}-4R_{\mu\nu}R^{\mu\nu}+R^{2}.\label{eq:GB combination}
\end{equation}
The Gauss-Bonnet term is topological in four dimensions and vanishes in fewer than four
dimensions.  Hence we expect the equations of motion to be independent of $\alpha_{\mathrm{\scriptscriptstyle GB}}$
for $d\le3$.  The
coefficients in \eqref{eq:hd-Lagrangian, new basis} and \eqref{eq:general action}
are related via 
\begin{eqnarray}
\alpha_{1} & = & \alpha_{\mathrm{{\scriptscriptstyle GB}}}+\alpha_{\mathrm{{\scriptscriptstyle W}}},\nonumber \\
\alpha_{2} & = & -4\alpha_{\mathrm{{\scriptscriptstyle GB}}}-\frac{4}{d-1}\alpha_{\mathrm{{\scriptscriptstyle W}}},\nonumber \\
\alpha_{3} & = & \alpha_{\mathrm{{\scriptscriptstyle GB}}}+\frac{2}{d(d-1)}\alpha_{\mathrm{{\scriptscriptstyle W}}}+\alpha_{\mathrm{{\scriptscriptstyle R}}}.\label{eq:new alpha's}
\end{eqnarray}
In this new basis, the final solution is given by the Lifshitz metric (\ref{eq:lifmet}),
the Maxwell field (\ref{eq:F Ansatz}), and the dilaton
\begin{equation}
\phi = -\frac{2(d-1)}{\lambda_{1}}r+C,
\end{equation}
where C is a constant of integration. The gauge kinetic function is $f(\phi)=e^{\lambda_1\phi}$, where
\begin{equation}
\lambda_{1}^{2}=\frac{(d-1)(z+d-1)}{Q^{2}e^{-\lambda_{1}C}}.\label{eq:Lifshitz solution, phi}
\end{equation}
The electric charge $Q$ and the cosmological constant $\Lambda$ are given in terms of $z$
according to
\begin{eqnarray}
Q^{2}e^{-\lambda_{1}C} & = & \frac{1}{2}(z-1)(z+d-1)\left[1-\frac{4(d-2)}{d}\alpha_{\mathrm{{\scriptscriptstyle W}}}z(z-d-1)-2(d-2)(d-3)\alpha_{\mathrm{{\scriptscriptstyle GB}}}\right]\nonumber \\
&  & -2\alpha_{\mathrm{{\scriptscriptstyle R}}}\bigg[z^{4}+2(d-\frac{3}{2})z^{3}+\frac{3}{2}(d-1)z^{2}+\frac{1}{2}(d-1)(d^{2}-4d+2)z-\frac{1}{2}d(d-1)^{2}\bigg],\nn\\
\label{eq:Lifshitz solution Q^2}\\
\Lambda & = & (z+d-1)(z+d-2)-\frac{4(d-2)^{2}}{d}\alpha_{\mathrm{{\scriptscriptstyle W}}}z(z-1)\left(z-2\frac{d-1}{d-2}\right)\nonumber \\
 &  & -2(d-2)(d-3)\alpha_{\mathrm{{\scriptscriptstyle GB}}}\left[z^{2}+2(d-\frac{3}{2})z+\frac{1}{2}(d-1)(d-4)\right]\nonumber \\
 &  & +4\alpha_{\mathrm{{\scriptscriptstyle R}}}\left[z^{3}-\frac{3}{2}(d-1)(d-\frac{8}{3})z^{2}-(d-1)(d^{2}-\frac{7}{2}d+2)z-\frac{1}{4}d(d-1)^{2}(d-4)\right].\nn\\
 \label{eq:Lifshitz solution, Lambda}
\end{eqnarray}
As expected, for $d=2$ and $3$, the Gauss-Bonnet combination does not contribute
to the equations of motion. Notice also that due to the shift symmetry
$\phi\mapsto\phi+C$, $F_{\text{\ensuremath{\mu\nu}}}\mapsto F_{\text{\ensuremath{\mu\nu}}}e^{-\frac{1}{2}\lambda_{1}C}$,
only the combination $Q^{2}e^{-\lambda_{1}C}$ is fixed. 

We see that the higher derivative action \eqref{eq:general action} admits Lifshitz
solutions with an electric background gauge potential and $\phi\propto r$.
The ``running'' of the dilaton has physical consequences: The effective
gauge coupling $f^{-\frac{1}{2}}$ runs from weak coupling in the
IR ($r\rightarrow-\infty$) to strong coupling in the UV  ($r\rightarrow\infty$)%
\footnote{In four dimensions, we can use electric-magnetic duality to obtain a
magnetic solution, $\widetilde{F}\equiv f\left(\phi\right)\ast F=Q_{m}dx\wedge dy$,
with magnetic charge $Q_{m}$. Since the duality transformation also
requires $f\mapsto f^{-1}$, the dilaton now runs towards strong coupling
in the IR.}. 

The above solution is the straightforward generalization of the previously
known Maxwell-dilaton background to the case of four-derivative gravity.
The effect of the higher derivative corrections is to renormalize
the cosmological constant and electric charge by inducing corrections
of order $z^{4}$. We will demonstrate below that this leads to some
nontrivial features of the solution.

\subsection{Lifshitz solutions in Einstein-Weyl gravity}

Let us now focus on the special case of Einstein-Weyl gravity. This
theory will be of particular interest to us in the following section,
where we will construct smooth flows from AdS$_{4}$ to Lifshitz to
AdS$_{2}\times\mathbb{R}^{2}$. Lifshitz solutions in pure Einstein-Weyl
gravity without additional matter fields have also been studied in
\cite{Lu:2012xu}. 

Setting $\alpha_{\mathrm{{\scriptscriptstyle GB}}}=\alpha_{\mathrm{{\scriptscriptstyle R}}}=0$,
the solution simplifies to
\begin{eqnarray}
Q^{2}e^{-\lambda_{1}C} & = & \frac{1}{2}(z-1)(z+d-1)\left(1-\frac{4(d-2)}{d}\alpha_{\mathrm{{\scriptscriptstyle W}}}z(z-d-1)\right),\label{eq:L soln, Q^2 for Weyl^2}\\
\Lambda & = & (z+d-1)(z+d-2)-\frac{4(d-2)^{2}}{d}\alpha_{\mathrm{{\scriptscriptstyle W}}}z(z-1)\left(z-2\frac{d-1}{d-2}\right).\label{eq:L soln, Lambda for Weyl^2}
\end{eqnarray}
This solution has some interesting features. For $d=2$, the Weyl-tensor
vanishes identically and so there are no higher derivative corrections.
Next, notice that if $Q^{2}\rightarrow0$, $\lambda_{1}\rightarrow\infty$,
$\phi\rightarrow\mathrm{const.}$, the matter fields decouple and
we recover a purely gravitational solution. There are two distinct
ways to achieve this: The first one is the case $z=1$, corresponding
to pure AdS$_{d+1}$ without matter fields. Note that because the
Weyl tensor vanishes in AdS, the cosmological constant is not renormalized. 

As a second possibility, we may choose 
\begin{equation}
\alpha_{\mathrm{{\scriptscriptstyle W}}}=\frac{d}{4(d-2)z(z-d-1)}.\label{eq:decoupling alpha}
\end{equation}
In this case we recover purely gravitational Lifshitz solutions, with
\begin{eqnarray}
Q^{2}e^{-\lambda_{1}C} & = & 0,\nonumber \\
\phi & = & \mathrm{const.},\nonumber \\
\Lambda & = & (z+d-1)(z+d-2)-(d-2)(z-1)\frac{z-2\frac{d-1}{d-2}}{z-d-1}.
\end{eqnarray}
It is interesting to consider the limit of conformal gravity, where
$\alpha_{\mathrm{{\scriptscriptstyle W}}}\rightarrow\infty$. From
\eqref{eq:decoupling alpha}, we expect the scaling parameter to take
two possible values, $z=0$, or $z=d+1$. However, in the latter
case, $\Lambda$ blows up for general $d$. It is only in the case
$d=3$ that the second solution with $z=4$ is well behaved. Finally,
notice also that for any given $\alpha$ and $\lambda_{1}$, there
may be multiple solutions for $z$ (see also
Appendix~\ref{sec:Lifshitz solutions in alternative coordinate system}).

\section{Smoothing out the singularity}

\label{sec:smoothing}The Lifshitz solutions of the previous section
have a physical singularity in the infrared. For $z\neq1$, an infalling
extended object, such as a string, experiences infinitely strong tidal
forces as $r\rightarrow-\infty$ \cite{Horowitz:2011gh}. Hence pure
Lifshitz solutions are `IR incomplete'. However, one might argue that
this kind of pathological behavior is simply a signal that our solutions
should not be trusted in this particular regime and the singularity
would presumably be resolved in a more complete string theory picture.
Some compelling evidence supporting this point of view has been presented
in \cite{Harrison:2012vy,Bhattacharya:2012zu,Bao:2012yt}.

The analysis of the previous section suggests a straightforward way
of resolving the Lifshitz singularity: In general, a nonzero coupling
of the dilaton to higher derivative terms will generate corrections
to its effective potential. In this section, we will use a simple
toy model in four dimensions to show that by choosing such a coupling
appropriately, the dilaton can be stabilized at some finite value
$\phi_{0}$. As a result, the geometry flows smoothly from Lifshitz
to AdS$_{2}\times\mathbb{R}^{2}$ in the deep IR, which is free of
physical singularities. 

In order to imitate the effect of generic higher derivative corrections
from string theory, we consider the following theory:
\begin{equation}
S=\int d^{4}x\sqrt{-g}\left(R+\Lambda-\tfrac{1}{2}\left(\partial\phi\right)^{2}-f\left(\phi\right)F_{\mu\nu}F^{\mu\nu}+g(\phi)C_{\mu\nu\rho\sigma}C^{\mu\nu\rho\sigma}\right).\label{eq: 4d action with f(phi), g(phi)}
\end{equation}
Since the Weyl tensor vanishes in AdS$_{4}$, the higher derivative
terms do not source the dilaton in the UV. We therefore expect a smooth
flow from AdS$_{4}$ to Lifshitz, much like the domain-wall solutions
found in \cite{Goldstein:2009cv, Perlmutter:2010qu}.  As we flow further towards the IR, the Weyl-squared
term ought to become more important, and the dilaton-Weyl coupling, $g(\phi)$,
may then stabilize the dilaton.  To be concrete, we choose $g(\phi)$ to be
\begin{equation}
g(\phi)=\frac{3}{4}(\alpha+\beta e^{\lambda_{2}\phi}).\label{eq:Ansatz for g(phi)}
\end{equation}
For $\beta e^{\lambda_{2}\phi}\ll\alpha$, $g(\phi)$ is approximately constant
and we expect to find Lifshitz scaling solutions of the form described
in the previous section. With an appropriate choice of parameters,
the exponential becomes more and more important as $\phi$ runs towards
weak coupling and it eventually stabilizes the dilaton in the deep
IR.

Since we have introduced a Weyl-squared correction, it is convenient to choose the following
parametrization of the metric\footnote{We will work in units where $L=1$ in what follows. }:
\begin{equation}
ds^{2}=a^{2}(r)\left(-dt^{2}+dr^{2}+b^{2}(r)(dx^{2}+dy^{2})\right),\label{eq:metric parametrization}
\end{equation}
With this choice, the Weyl-invariance of the higher derivative
Lagrangian is manifest as a rescaling of $a(r)$. In practice, this
means that only $b(r)$ will receive higher derivative corrections
in the equations of motion. Fixing $\Lambda=1$, the AdS$_{4}$ solution
is given by
\begin{equation}
a=\frac{\sqrt{6}}{r},\qquad b=\mathrm{const.},
\end{equation}
while the Lifshitz solution takes the form 
\begin{equation}
a\propto\frac{1}{r},\qquad b\propto r^{\tilde{z}}.\label{eq:Lifshitz solns in new gauge}
\end{equation}
For this metric, the scaling symmetry \eqref{eq:Lifshitz scaling symmetry} and \eqref{eq:Lss2}
becomes
\begin{equation}
t\rightarrow\lambda t,\qquad x\rightarrow\lambda^{1-\tilde{z}}x,\qquad r\rightarrow\lambda r.
\end{equation}
In changing from the more common form of the metric, \eqref{eq:lifbulk}, to the Weyl form,
\eqref{eq:metric parametrization},
we need to make the following identifications:
\begin{eqnarray}
z & = & \frac{1}{1-\tilde{z}},\nonumber \\
L & \rightarrow & \frac{L}{1-\tilde{z}},\nonumber \\
\alpha & \rightarrow & (1-\tilde{z})^{2}\alpha.
\end{eqnarray}
As before, we choose a background electric charge:
\begin{equation}
F=\frac{Q}{b^{2}f(\phi)}dr\wedge dt.
\end{equation}
Einstein's equations are:
\begin{eqnarray}
T_{00} & = & -2\left(\frac{a^{\prime\prime}}{a}+\frac{b^{\prime\prime}}{b}\right)+\left(\frac{a^{\prime}}{a}\right)^{2}-\left(\frac{b^{\prime}}{b}\right)^{2}-4\frac{a^{\prime}b^{\prime}}{ab}\nonumber \\
&  & -\frac{4}{3}\frac{g(\phi)}{a^{2}}\Biggl[\frac{b^{\left(4\right)}}{b}+\frac{b^{\left(3\right)}b^{\prime}}{b^{2}}-2\frac{b^{\prime\prime}\left(b^{\prime}\right)^{2}}{b^{3}}-\frac{1}{2}\left(\frac{b^{\prime\prime}}{b}\right)^{2}+\frac{1}{2}\left(\frac{b^{\prime}}{b}\right)^{4}+\left(\frac{b^{\prime\prime}}{b}-\left(\frac{b^{\prime}}{b}\right)^{2}\right)\frac{g^{\prime}}{g}\phi^{\prime\prime}\nonumber \\
&  &\kern4em+ \left(2\frac{b^{\left(3\right)}}{b}-\frac{b^{\prime}b^{\prime\prime}}{b^{2}}-\left(\frac{b^{\prime}}{b}\right)^{3}\right)\frac{g^{\prime}}{g}\phi^{\prime}+\left(\frac{b^{\prime\prime}}{b}-\left(\frac{b^{\prime}}{b}\right)^{2}\right)\frac{g^{\prime\prime}}{g}\left(\phi^{\prime}\right)^{2}\Biggr],\label{eq:T00}\\
T_{rr} & = & 3\left(\frac{a^{\prime}}{a}\right)^{2}+\left(\frac{b^{\prime}}{b}\right)^{2}+4\frac{a^{\prime}b^{\prime}}{ab}\nonumber \\
 &  & +\frac{4}{3}\frac{g\left(\phi\right)}{a^{2}}\left[-\frac{b^{\left(3\right)}b^{\prime}}{b^{2}}+\frac{1}{2}\left(\frac{b^{\prime}}{b}\right)^{4}+\frac{1}{2}\left(\frac{b^{\prime\prime}}{b}\right)^{2}+\left(\left(\frac{b^{\prime}}{b}\right)^{3}-\frac{b^{\prime}b^{\prime\prime}}{b^{2}}\right)\frac{g^{\prime}}{g}\phi^{\prime}\right],\label{eq:Trr}\\
\frac{T_{ii}}{b^{2}} & = & 2\frac{a^{\prime\prime}}{a}+\frac{b^{\prime\prime}}{b}-\left(\frac{a^{\prime}}{a}\right)^{2}+2\frac{a^{\prime}b^{\prime}}{ab}\nonumber \\
&  & -\frac{4}{3}\frac{g(\phi)}{a^{2}}\Biggl[\frac{1}{2}\frac{b^{\left(4\right)}}{b}+\frac{1}{2}\left(\frac{b^{\prime}}{b}\right)^{4}-\frac{\left(b^{\prime}\right)^{2}b^{\prime\prime}}{b^{3}}+\frac{1}{2}\left(\frac{b^{\prime\prime}}{b}-\left(\frac{b^{\prime}}{b}\right)^{2}\right)\left(\frac{g^{\prime}}{g}\phi^{\prime\prime}+\frac{g^{\prime\prime}}{g}\left(\phi^{\prime}\right)^{2}\right)\nonumber \\
&  &\kern4em +\left(\frac{b^{\left(3\right)}}{b}-\frac{b^{\prime}b^{\prime\prime}}{b^{2}}\right)\frac{g^{\prime}}{g}\phi^{\prime}\Biggr],\label{eq:Tii}
\end{eqnarray}
with
\begin{eqnarray}
T_{00} & = & \frac{Q^{2}}{a^{2}b^{4}f\left(\phi\right)}+\frac{1}{4}\left(\phi^{\prime}\right)^{2}-\frac{a^{2}\Lambda}{2},\\
T_{rr} & = & -\frac{Q^{2}}{a^{2}b^{4}f\left(\phi\right)}+\frac{1}{4}\left(\phi^{\prime}\right)^{2}+\frac{a^{2}\Lambda}{2},\\
\frac{T_{ij}}{b^{2}} & = & \delta_{ij}\left(\frac{Q^{2}}{a^{2}b^{4}f\left(\phi\right)}-\frac{1}{4}\left(\phi^{\prime}\right)^{2}+\frac{a^{2}\Lambda}{2}\right).
\end{eqnarray}
If we demand that $\phi$ depends only on $r$, the dilaton equation
of motion simplifies to
\begin{equation}
\phi^{\prime\prime}+2\left(\frac{a^{\prime}}{a}+\frac{b^{\prime}}{b}\right)\phi^{\prime}+a^{2}V_{\mathrm{eff}}^{\prime}(\phi)=0,\label{eq:phi eom}
\end{equation}
where
\begin{equation}
V_{\mathrm{eff}}^{\prime}(\phi)\equiv\frac{2Q^{2}}{a^{4}b^{4}}\frac{f^{\prime}(\phi)}{f^{2}(\phi)}+\frac{4}{3a^{4}}\bigg(\frac{\mathrm{d}^{2}\log(b)}{\mathrm{d}r^{2}}\bigg)^{2}g^{\prime}(\phi).
\end{equation}
Hence the effect of the higher derivative terms is to generate a correction
to the effective dilaton potential.

We would like to find out which
choices of $g(\phi)$ allow for an emerging AdS$_{2}\times\mathbb{R}^{2}$
geometry in the deep IR. Corresponding to AdS$_2\times\mathbb{R}^2$, we make the ansatz
\begin{eqnarray}
a(r) & = & \frac{1}{r},\nonumber \\
b(r) & = & b_{0}r,\nonumber \\
\phi(r) & = & \phi_{0}.
\label{eq:AdS2soln}
\end{eqnarray}
Solving \eqref{eq:T00}-\eqref{eq:Tii}, we find 
\begin{eqnarray}
\Lambda & = & 1,\nonumber \\
\frac{Q^{2}}{b_{0}^{4}} & = & \frac{f(\phi_{0})}{2}\left(1-\frac{4}{3}g(\phi_{0})\right).\label{eq:Q^2 condition}
\end{eqnarray}
Since only the ratio ${Q}/{b_{0}^{2}}$ is fixed, we are free
to set $b_{0}\equiv1$ in what follows. Equation \eqref{eq:phi eom}
gives us the condition
\begin{equation}
V_{\mathrm{eff}}^{\prime}(\phi_{0})=\frac{f^{\prime}(\phi_{0})}{f(\phi_{0})}\left(1-\frac{4}{3}g(\phi_{0})\right)+\frac{4}{3}g^{\prime}(\phi_{0})=0.\label{eq:local minimum condition}
\end{equation}
Let us now specialize to the case $f\left(\phi\right)=e^{\lambda_{1}\phi}$.
Since the dilaton runs towards weak coupling as $r\rightarrow\infty$,
this ansatz is valid even in the deep IR. With our choice of $g\left(\phi\right)$,
the solution to \eqref{eq:Q^2 condition} and \eqref{eq:local minimum condition}
is given by
\begin{eqnarray}
Q^{2} & = & \frac{(\alpha-1)\lambda_{2}}{\lambda_{1}-\lambda_{2}}\nonumber, \\
\phi_{0} & = & \frac{1}{\lambda_{2}}\log\left(\frac{\lambda_{1}}{\lambda_{1}-\lambda_{2}}\frac{1-\alpha}{\beta}\right).\label{eq:IR value of dilaton}
\end{eqnarray}
Clearly this solution only makes sense for a certain choice of $\lambda_{i}$,
$\alpha$, $\beta$. We will discuss the constraints on these parameters
at the end of the next section.

\subsection{Perturbations around $\mathrm{AdS}_{2}\times\mathbb{R}^{2}$}

We would like to find numerical solutions that smoothly interpolate
between AdS$_{4}$ and AdS$_{2}\times\mathbb{R}^{2}$, with some intermediate
Lifshitz regime. This is most easily accomplished numerically by using
the ``shooting'' technique, starting in the deep IR ($r\rightarrow\infty$).
The initial conditions have to be chosen such that we follow perturbations
that are \textit{irrelevant} in the IR. These are perturbations that
fall off faster than the background solution as $r\rightarrow\infty$.
In other words, they allow a smooth flow \textit{away from} AdS$_{2}\times\mathbb{R}^{2}$
as $r$ decreases. Requiring the existence of such perturbations will
introduce nontrivial constraints on the parameters of our model.

We start by perturbing the AdS$_{2}\times\mathbb{R}^{2}$ solution, (\ref{eq:AdS2soln}),
in the following way
\begin{equation}
a\left(r\right)=\frac{1}{r}+\delta a\left(r\right),\qquad b\left(r\right)=r+\delta b\left(r\right),\qquad\phi\left(r\right)=\phi_{0}+\delta\phi\left(r\right).
\end{equation}
Using the conditions \eqref{eq:Q^2 condition} and \eqref{eq:local minimum condition}
repeatedly, the linearized equations of motion may be written as
\begin{eqnarray}
\frac{3}{2g_{0}}\frac{\left(r^{3}\delta a^{\prime}\right)^{\prime}}{r^{2}}+\left(r\delta b^{\left(3\right)}\right)^{\prime}-2\frac{f_{0}g_{0}^{\prime}}{f_{0}^{\prime}g_{0}}\left(\frac{\delta b^{\prime}}{r}\right)^{\prime}-\frac{g_{0}^{\prime}}{g_{0}}\frac{\left(r\delta\phi^{\prime}\right)^{\prime}}{r}+\frac{g_{0}^{\prime}}{g_{0}}\frac{\delta\phi}{r^{2}} & = & 0,\qquad\\
\frac{3}{2g_{0}}\left(r^{2}\delta a\right)^{\prime}+r^{2}\delta b^{\left(3\right)}-2\frac{f_{0}g_{0}^{\prime}}{f_{0}^{\prime}g_{0}}\delta b^{\prime}-\frac{g_{0}^{\prime}}{g_{0}}\left(r\delta\phi\right)^{\prime} & = & 0,\\
-3\left(r^{2}\delta a^{\prime}\right)^{\prime}+r^{2}g_{0}\delta b^{\left(4\right)}-2\left(g_{0}+\frac{3}{4}\right)r^{2}\left(\frac{\delta b^{\prime}}{r^{2}}\right)^{\prime}-6\frac{\delta b}{r^{2}}-g_{0}^{\prime}r\left(\delta\phi^{\prime\prime}-2\frac{\delta\phi}{r^{2}}\right) & = & 0,\\
\delta\phi^{\prime\prime}+V_{\mathrm{eff}}^{\prime\prime}\left(\phi_{0}\right)\frac{\delta\phi}{r^{2}}-\frac{8}{3}g_{0}^{\prime}r\left(\frac{\delta b^{\prime}}{r^{2}}\right)^{\prime} & = & 0,\label{eq:dilaton perturbation eom}
\end{eqnarray}
where $f_{0}\equiv f\left(\phi_{0}\right)$, etc. The presence of
the $\delta b$ term in the last equation emphasizes the fact that
the higher derivative corrections generate a \textit{gravitational}
effective potential for the dilaton. This is different from the case
of a quantum-corrected $f\left(\phi\right)$, and will in general
lead to a nontrivial mixing of $\phi$ perturbations with gravitational
perturbations. Since the first three equations are related via a
Bianchi identity, it is possible to eliminate the $\delta b^{\left(4\right)}$
terms and reduce the system to a third order coupled ODE. Hence there
are only seven independent solutions:
\begin{eqnarray}
&&\delta a  =  -r,\kern5.7em
\delta b  =  r^{3},\kern4.2em
\delta\phi  =  0; \\
&&\delta a  =  -\frac{\frac{3}{4}+\log(r)}{r^{2}},\qquad
\delta b  =  \log r,\kern3.6em
\delta\phi  =  \frac{\xi}{r};\label{eq:irrelevant gravitational perturbations}
\\
&&\delta a  =  -\frac{1}{r^{2}},\kern5em
\delta b  =  1,\kern4.6em
\delta\phi = 0;\label{eq:irrelevant gravitational perturbation 2}
\\
&&\delta a  =  A_{0}r^{\nu-1},\kern3.8em
\delta b  =  B_{0}r^{\nu+1},\qquad
\delta\phi  =  P_{0}r^{\nu}.\label{eq:dilaton perturbations}
\end{eqnarray}
Here
\begin{equation}
\xi = \frac{6\lambda_{1}\lambda_{2}\left(1-\alpha\right)}{\lambda_{1}\lambda_{2}\left(\lambda_{1}+\lambda_{2}\right)\left(\alpha-1\right)+2\left(\lambda_{2}-\lambda_{1}\right)},
\end{equation}
and the constants $A_0$, $B_0$ and $P_0$ in (\ref{eq:dilaton perturbations}) are related by
\begin{eqnarray}
A_{0} & = & \frac{2g_{0}}{3}\left(\frac{g_{0}^{\prime}}{g_{0}}\left(P_{0}+2\frac{f_{0}}{f_{0}^{\prime}}\right)-\nu\left(\nu-1\right)\right)B_{0},\nn\\
P_{0} & = & \frac{8}{3}\frac{g_{0}^{\prime}\left(\nu+1\right)\left(\nu-2\right)}{V^{\prime\prime}\left(\phi_{0}\right)+\nu\left(\nu-1\right)}B_{0}.
\end{eqnarray}
There are four solutions for the exponent in \eqref{eq:dilaton perturbations}:
\begin{eqnarray}
\nu & \equiv & \frac{1}{2}+\widetilde{\nu},\nonumber \\
2\widetilde{\nu}^{2} & = & -\left(V_{\mathrm{eff}}^{\prime\prime}\left(\phi_{0}\right)-\frac{1}{4}-\frac{8}{3}\frac{\left(g_{0}^{\prime}\right)^{2}}{g_{0}}+x\right)\pm\left[\left(V_{\mathrm{eff}}^{\prime\prime}\left(\phi_{0}\right)-\frac{1}{4}-\frac{8}{3}\frac{\left(g_{0}^{\prime}\right)^{2}}{g_{0}}-x\right)^{2}-\frac{16}{3}\left(\frac{g_{0}^{\prime}}{g_{0}}\right)^{2}\right]^{\frac{1}{2}}\!\!,\nonumber \\
x & \equiv & \frac{5}{12}-\frac{1}{2g_{0}}-\frac{4}{3}\frac{f_{0}g_{0}^{\prime}}{f_{0}^{\prime}g_{0}}.
\end{eqnarray}
For our choice of $g\left(\phi\right)$, given by (\ref{eq:Ansatz for g(phi)}), we get 
\begin{eqnarray}
\widetilde{\nu}&=&\pm\frac{1}{2}\Biggl[1-\frac{\left(1-\alpha\right)\lambda_{2}}{1-\alpha\frac{\lambda_{2}}{\lambda_{1}}}\biggl[2\lambda_{1}-\frac{4}{3\lambda_{1}}+2\alpha\lambda_{2}
\nn\\
&&\kern9em\pm\frac{2}{\lambda_{1}}\left(\lambda_{1}^{4}+2\alpha\lambda_{2}\lambda_{1}^{3}+\left(\alpha^{2}\lambda_{2}^{2}-4\right)\lambda_{1}^{2}+\frac{4}{3}\alpha\lambda_{1}\lambda_{2}+\frac{4}{9}\right)^{\frac{1}{2}}\biggr]\Biggr]^{\frac{1}{2}}\!\!.\qquad\label{eq:Ctilde solution}
\end{eqnarray}

Regardless of the form of the effective potential, there always exist
two irrelevant perturbations, \eqref{eq:irrelevant gravitational perturbations}
and \eqref{eq:irrelevant gravitational perturbation 2}. Whether or
not the solutions \eqref{eq:dilaton perturbations} are irrelevant
depends on the choice of parameters. Although in general there is
a mixing of $\phi$ with $a$ and $b$ due to the dilaton coupling
to $C_{\mu\nu\rho\sigma}^2$, one can check that for $g\left(\phi\right)\equiv0$
the ansatz \eqref{eq:dilaton perturbations} reproduces the purely
dilatonic perturbations of the two-derivative theory \cite{Harrison:2012vy}.
Although not technically correct, we will therefore still refer to
those perturbations as ``dilaton perturbations'' in what follows.

\begin{figure}[tp]
\centering
\includegraphics[width=7.6cm]{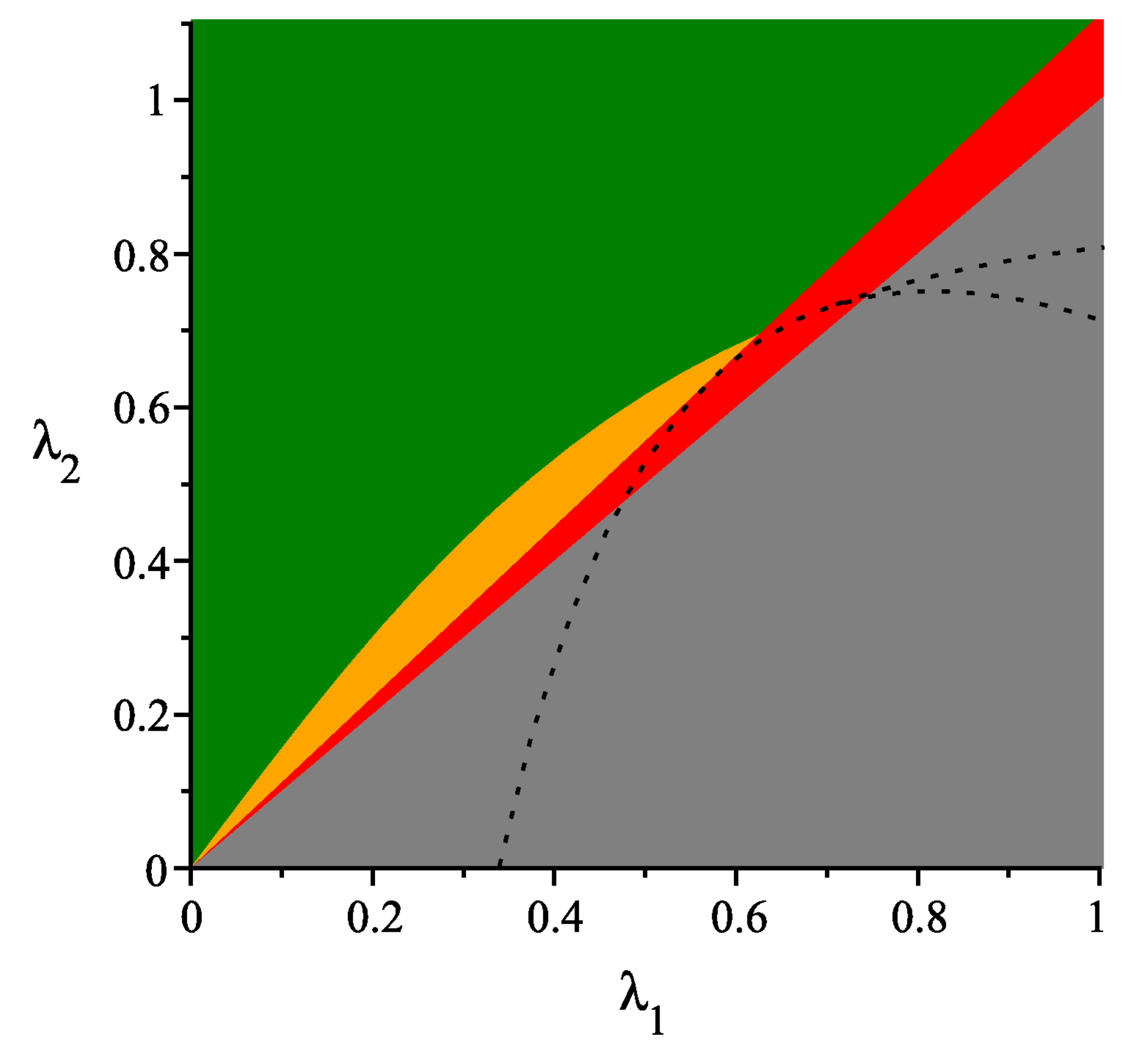}
\hfill
\includegraphics[width=7.8cm]{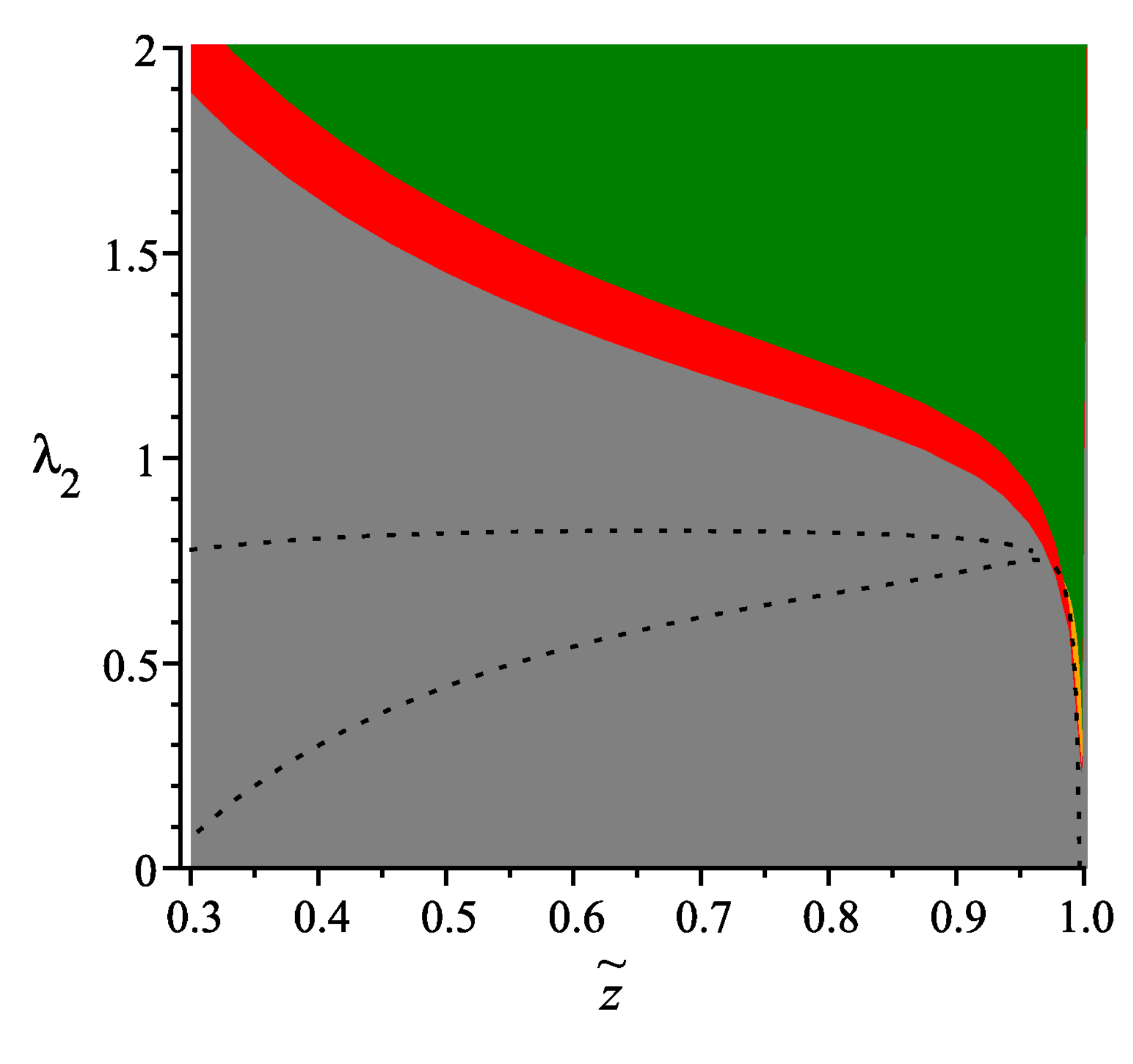}
\caption{\label{fig:contours1}Plot of different regions in parameter space,
characterized by the number of irrelevant dilaton perturbations ($\alpha=0.9$).
The regions are bounded by the curves \eqref{eq:lambda2 curve1},
\eqref{eq:lambda2 curve2},  \eqref{eq:two different cases for alpha (<>1)} and $\lambda_2=\lambda_1$. They are colored as follows: $g\left(\phi\right)<0$
(red), $g\left(\phi\right)>0$ but no irrelevant perturbations or
oscillating modes (gold), $g\left(\phi\right)>0$ and at least one
irrelevant perturbation (green). In the gray area, at least one of
the conditions 1-4 is violated.}
\end{figure}

\begin{figure}[tp]
\centering
\includegraphics[width=7.8cm]{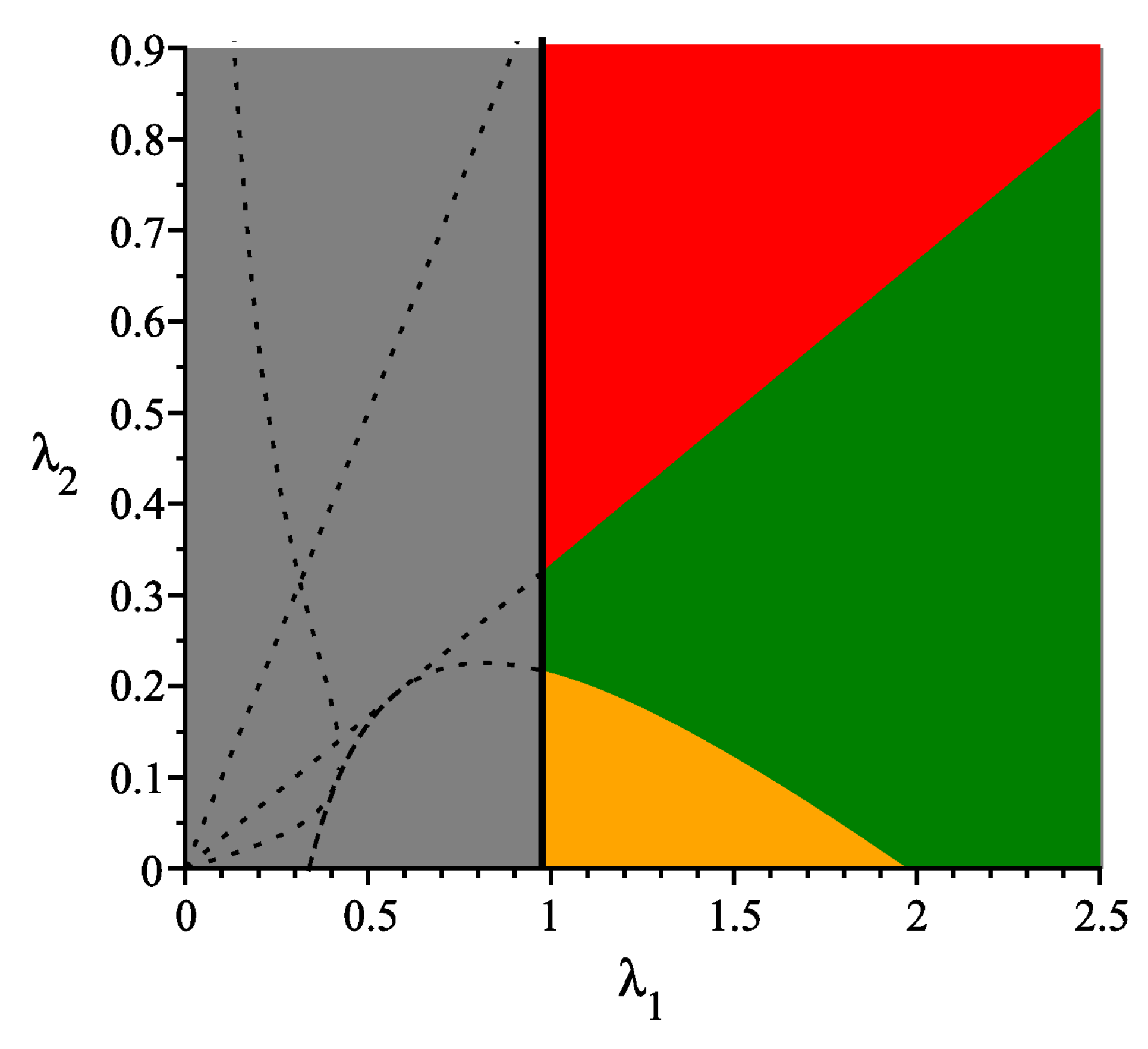}
\hfill
\includegraphics[width=7.6cm]{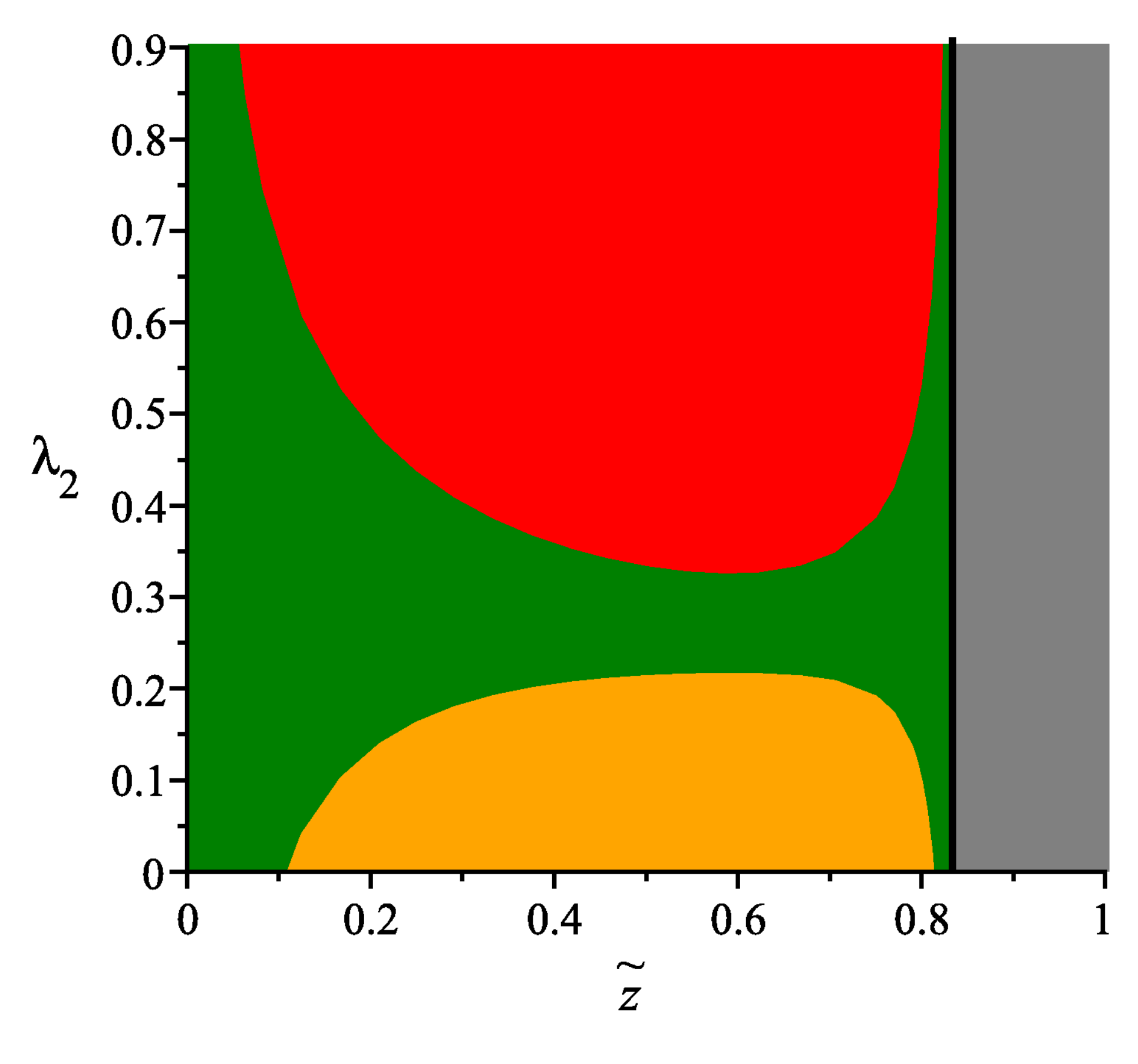}
\caption{\label{fig:contours2}The same plot for $\alpha=3$. Now the green
region has 2 irrelevant perturbations. There is a lower bound on $\lambda_{1}$
and an upper bound on $\tilde{z}$, as indicated by vertical lines.}
\end{figure}

To find the desired numerical solutions, we impose the following set
of conditions:
\begin{enumerate}
\item ${\lambda_{2}}/{\lambda_{1}}>0$: This ensures that $g\left(\phi\right)\approx\mathrm{const.}$
during the Lifshitz scaling stage and in the deep UV. Thus $g^{\prime}\left(\phi\right)$
only becomes important in the IR, where it stabilizes the dilaton.
Since \eqref{eq: 4d action with f(phi), g(phi)} is invariant under
$\phi\mapsto-\phi$, $\lambda_{i}\mapsto-\lambda_{i}$, we shall assume
without loss of generality that $\lambda_{1}>0$ and $\lambda_{2}>0$.
\item $V_{\mathrm{eff}}^{\prime}\left(\phi_{0}\right)=0$ for some $\phi_{0}$
(see \eqref{eq:local minimum condition}): The effective potential
stabilizes the dilaton and admits an AdS$_{2}\times\mathbb{R}^{2}$
solution.
\item We focus on the case $g\left(\phi_{0}\right)>0$. For negative $g(\phi)$,
we numerically find either singular solutions or solutions with $\phi^{\prime}\ll0$
as we approach AdS$_{4}$. It is unclear whether the sign of higher
derivative terms has a physical interpretation in terms of unitarity or causality or
a generalized null energy condition. 
\item $Q^{2}>0$, i.e. the vector potential is real-valued.
\item Our numerical analysis, as well as the analysis performed in \cite{Harrison:2012vy,Bhattacharya:2012zu}
strongly suggest that we need at least one of the dilaton perturbations
to be irrelevant in order to ``kick'' $\phi$ out of its local minimum
in the IR and roll towards large negative values in the UV. We therefore
demand that $\nu<0$ for at least one of the dilaton perturbations.
Notice that there can be at most two solutions that satisfy this condition.
\item We require $\nu$ to be real-valued; that is, we exclude oscillating
perturbations. We take the existence of complex eigenvalues as an
indication of a dynamical instability. However, due to the higher-derivative
nature of our theory, a more detailed analysis of the time-dependent
perturbations would be needed in order to determine whether the theory
is truly unstable for complex exponents.
\end{enumerate}
Let us now find out what these conditions imply for our parameters
$\alpha,\beta,\lambda_{1},\lambda_{2}$. Conditions 3 and 4 allow
for two possible choices:
\begin{eqnarray}
&&1)\quad\alpha\geq1,\kern3.8em\alpha\frac{\lambda_{2}}{\lambda_{1}}\leq1;\nn\\
&&2)\quad0<\alpha<1,\qquad\alpha\frac{\lambda_{2}}{\lambda_{1}}>1.\label{eq:two different cases for alpha (<>1)}
\end{eqnarray}
In both cases, condition 2 then requires that $\beta<0$. Recall that
in the electric case $\phi\leq\phi_{0}$, so choosing the sign of
$g(\phi)$ in the IR determines the sign everywhere%
\footnote{Although we will not consider the case of negative $g(\phi)$, let us point
out that in this case we would also have to take $\beta<0$ to satisfy
condition 2, so this is a universal result.}.
Finally, we would like the dilaton perturbations to be non-oscillating (condition
5) and demand that at least one of them should be irrelevant
(condition 6). The details of the corresponding calculations can be found
in Appendix~\ref{sec:Irrelevant Perturbations}. Our results are summarized
in Figures~\ref{fig:contours1} and \ref{fig:contours2}. In the green
region, all of our conditions are satisfied. The gray region is inconsistent
with conditions 1-4, while in the red region $g\left(\phi\right)<0$.
The yellow region has $g\left(\phi\right)>0$, but has either no irrelevant
dilaton perturbations, or oscillating modes.

For $\alpha<1$, we find either one or no irrelevant dilaton perturbations,
while for $\alpha>1$ we find either two or none. This result seems
to be related to the fact that in the $\alpha>1$ case, $\lambda_{1}^{2}\left(\tilde{z}\right)$
is not injective, i.e. there exist two possible scaling parameters
$\tilde{z}_{1},\tilde{z}_{2}$ for any given $\lambda_{1}$ (see
Appendix~\ref{sec:Lifshitz solutions in alternative coordinate system}). We
will address this issue further at the end of the following section.
Notice also that while for $\alpha<1$, all values of $\lambda_{1}$
and $\tilde{z}$ are allowed, for $\alpha>1$ there is a lower
bound on $\lambda_{1}$ and an upper bound on $\tilde{z}$. These
bounds stems from the condition that $\lambda_{1}^{2}>0$ and equation
\eqref{eq:l1^2(ztilde)}.
We conclude that for a given choice of $\alpha$, there exists a large
region in parameter space that is consistent with our conditions and
hence admits the desired AdS$_{4}\rightarrow \mathrm{Lif}_{4}^{z}
\rightarrow \mathrm{AdS}_{2}\times\mathbb{R}^{2}$ solutions.

\subsection{Numerical results}

In order to find numerical solutions to our equations, we proceed
as follows: We set initial conditions at large $r$ by adding irrelevant
perturbations to the exact AdS$_{2}\times\mathbb{R}^{2}$ solution:
\begin{eqnarray}
a\left(r\right) & = & \frac{1}{r}+\sum_{i=1}^{3(4)}D_{i}\delta a_{i},\nonumber \\
b\left(r\right) & = & r+\sum_{i=1}^{3(4)}D_{i}\delta b_{i},\nonumber \\
\phi\left(r\right) & = & \phi_{0}+\sum_{i=1}^{3(4)}D_{i}\delta\phi_{i}.\label{eq:Ansatz for initial conditions}
\end{eqnarray}
We focus here on the case $\alpha<1$, for which there are three irrelevant
perturbations. The case $\alpha>1$ is discussed briefly at the end
of this section. The amplitudes $D_{i}$ have to be tuned in order
to find a solution that has both an intermediate Lifshitz regime and
a smooth flow to AdS$_{4}$ in the UV. In practice, it is however
easier to choose a different basis for the perturbations in \eqref{eq:Ansatz for initial conditions},
which allows us to specify $a^{\prime},b^{\prime},\phi^{\prime}$
directly. The role of the three initial conditions is then roughly the
following: The value of $\phi^{\prime}$ determines how long the solution
stays approximately AdS$_{2}\times\mathbb{R}^{2}$. There is a minimum
value $\phi_{\mathrm{min}}^{\prime}$ that is required to ``kick''
$\phi$ out of its local minimum and run logarithmically during the
Lifshitz stage. The transition stage from AdS$_{2}\times\mathbb{R}^{2}$
to Lifshitz is shifted towards the IR as we increase $\phi^{\prime}$.
The value of $a^{\prime}$ determines the duration of the scaling
stage: We find that in the space of initial conditions, there exists
a two-dimensional submanifold $\left(a_{\mathrm{crit}}^{\prime}\left(\phi^{\prime}\right),b^{\prime},\phi^{\prime}\right)$
with attractor-like behavior. As we approach this critical plane,
we observe the emergence of an intermediate Lifshitz stage, which
gets wider and wider as $a^{\prime}$ approaches $a_{\mathrm{crit}}^{\prime}$
from below, while for $a^{\prime}>a_{\mathrm{crit}}^{\prime}$ the
solution becomes singular. Therefore, by tuning $a^{\prime}$ we can
in principle make the Lifshitz stage arbitrarily long. Finally, the
value of $b^{\prime}$ needs to be tuned in order to achieve a smooth
flow to AdS$_{4}$ in the UV.

\begin{table}[tp]
\centering
\begin{tabular}{|c|c|c|}
\hline 
\# & 1 & 2\tabularnewline
\hline 
$\lambda_{1}$ & $1.2$ & $1.1$\tabularnewline
\hline 
$\lambda_{2}$ & $2$ & $0.24$\tabularnewline
\hline 
$\alpha$ & $0.9$ & $3$\tabularnewline
\hline 
$\beta$ & $-1$ & $-1$\tabularnewline
\hline 
$Q$ & $0.20$ & $4.55$\tabularnewline
\hline 
$\phi_{0}$ & $-0.95$ & $3.91$\tabularnewline
\hline 
$a^{\prime}$ & $-4.9\cdot10^{-9}$ & $-1.5\cdot10^{-7}$\tabularnewline
\hline 
$b^{\prime}$ & $-5.8\cdot10^{-6}$ & $2.0\cdot10^{-4}$\tabularnewline
\hline 
$\phi^{\prime}$ & $8\cdot10^{-7}$ & $10^{-5}$\tabularnewline
\hline 
$b^{\prime\prime}$ & - & $10^{-5}$\tabularnewline
\hline 
$\tilde{z}$ & $0.73$ & $0.78$\tabularnewline
\hline 
$K$ & $0.91$ & $0.82$\tabularnewline
\hline 
$\phi_{\mathrm{{\scriptscriptstyle UV}}}$ & $-19.8$ & $-7.6$\tabularnewline
\hline 
$b_{\mathrm{{\scriptscriptstyle UV}}}$ & $1.5\cdot10^{-11}$ & $2.9\cdot10^{-8}$\tabularnewline
\hline 
\end{tabular}
\caption{\label{tab: initial conditions and parameters}Parameters, initial
conditions and fit parameters for numerical solutions.}
\end{table}

\begin{figure}[tp]
\centering
\includegraphics[clip,height=6cm]{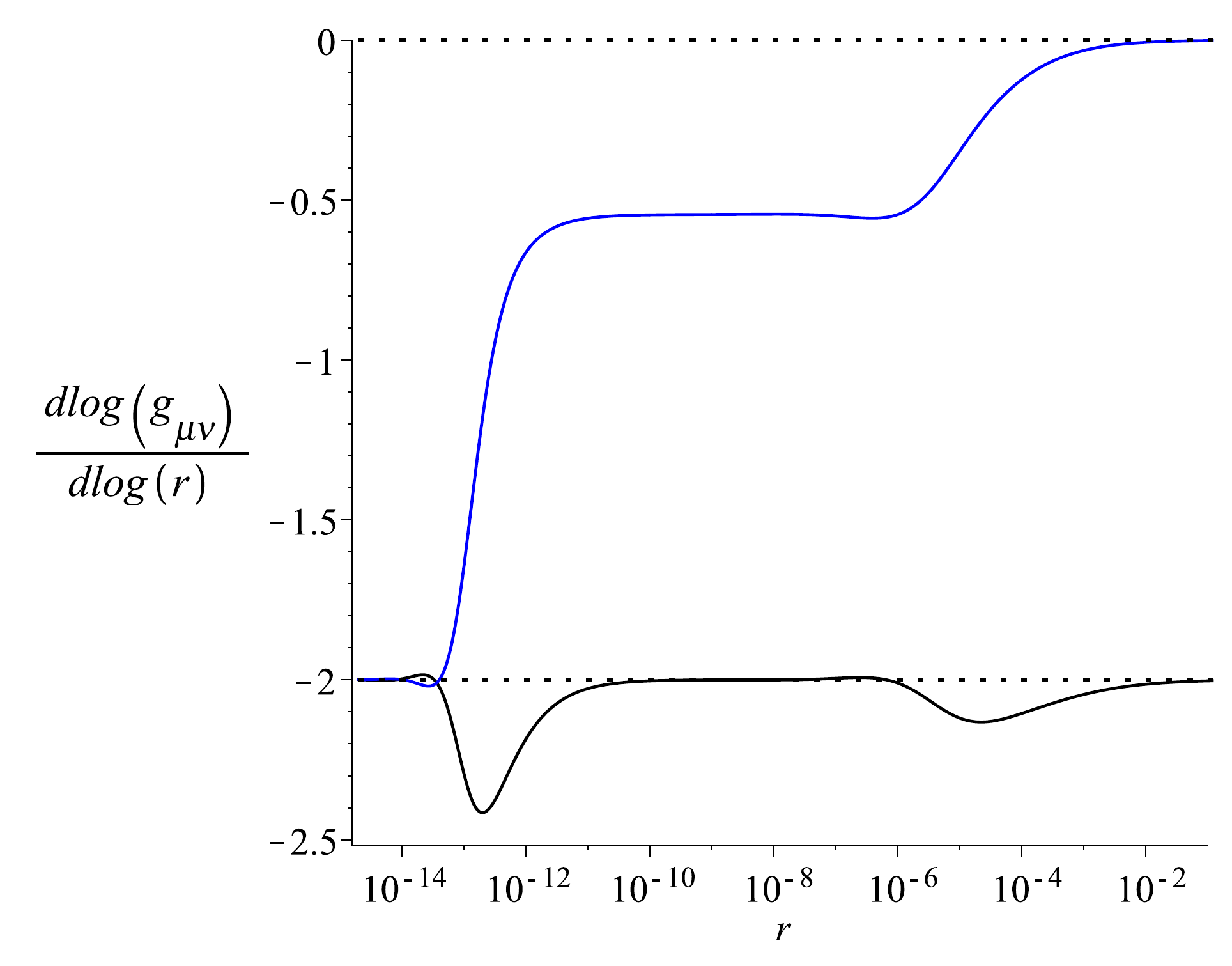}
\hfill
\includegraphics[clip,height=6cm]{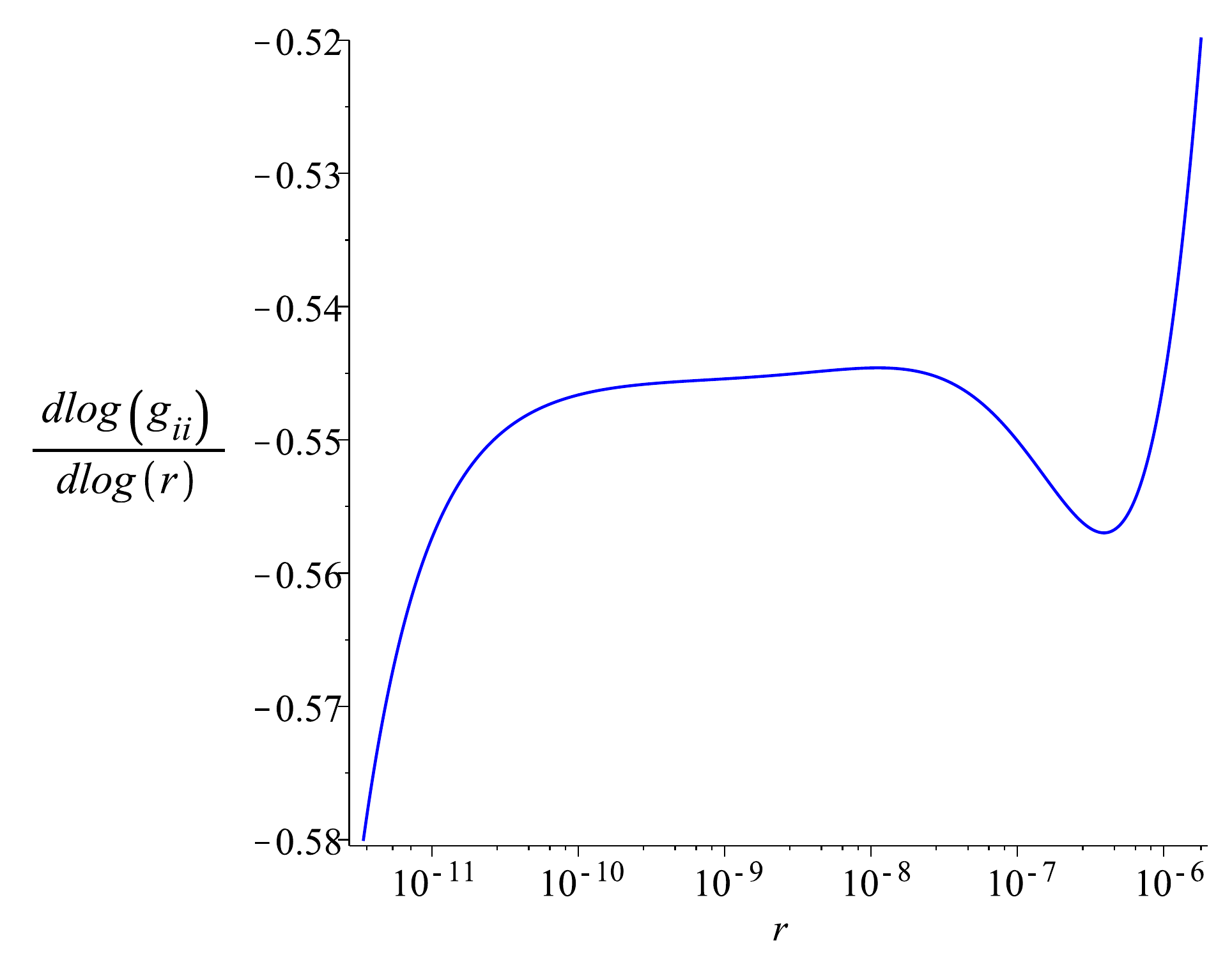}
\caption{\label{fig:g00-gxx}Plot of the metric components $g_{00}$ (black)
and $g_{ii}$ (blue) for solution \#1 (see Table~\ref{tab: initial conditions and parameters}).
The figure on the right is a magnified view of the Lifshitz region for $g_{ii}$.
Constant values of ${d\log g_{\mu\nu}}/{d\log r}$ indicate a
power-law relation. One can clearly see the emergence of an intermediate
Lifshitz geometry with $g_{00}\propto r^{-2}$ and $g_{ii}\propto r^{2\left(\tilde{z}-1\right)}$.
The dotted lines indicate the exact AdS$_{2}\times\mathbb{R}^{2}$
solution with $g_{ii}\propto r^{0}$ in the IR and AdS$_{4}$ with
$g_{ii}\propto r^{-2}$ in the UV.}
\end{figure}

\begin{figure}[tp]
\centering
\includegraphics[height=6cm]{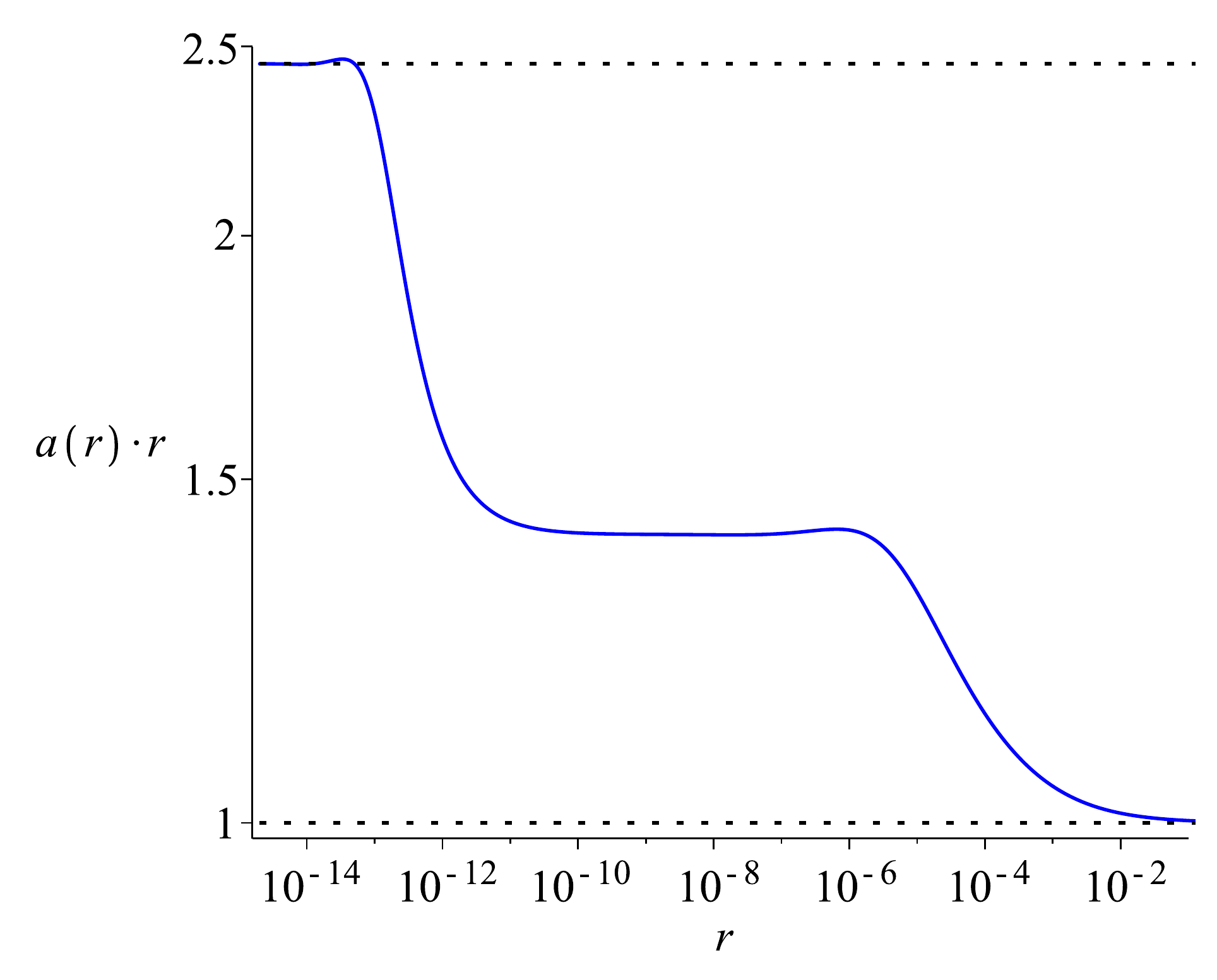}
\hfill
\includegraphics[height=6cm]{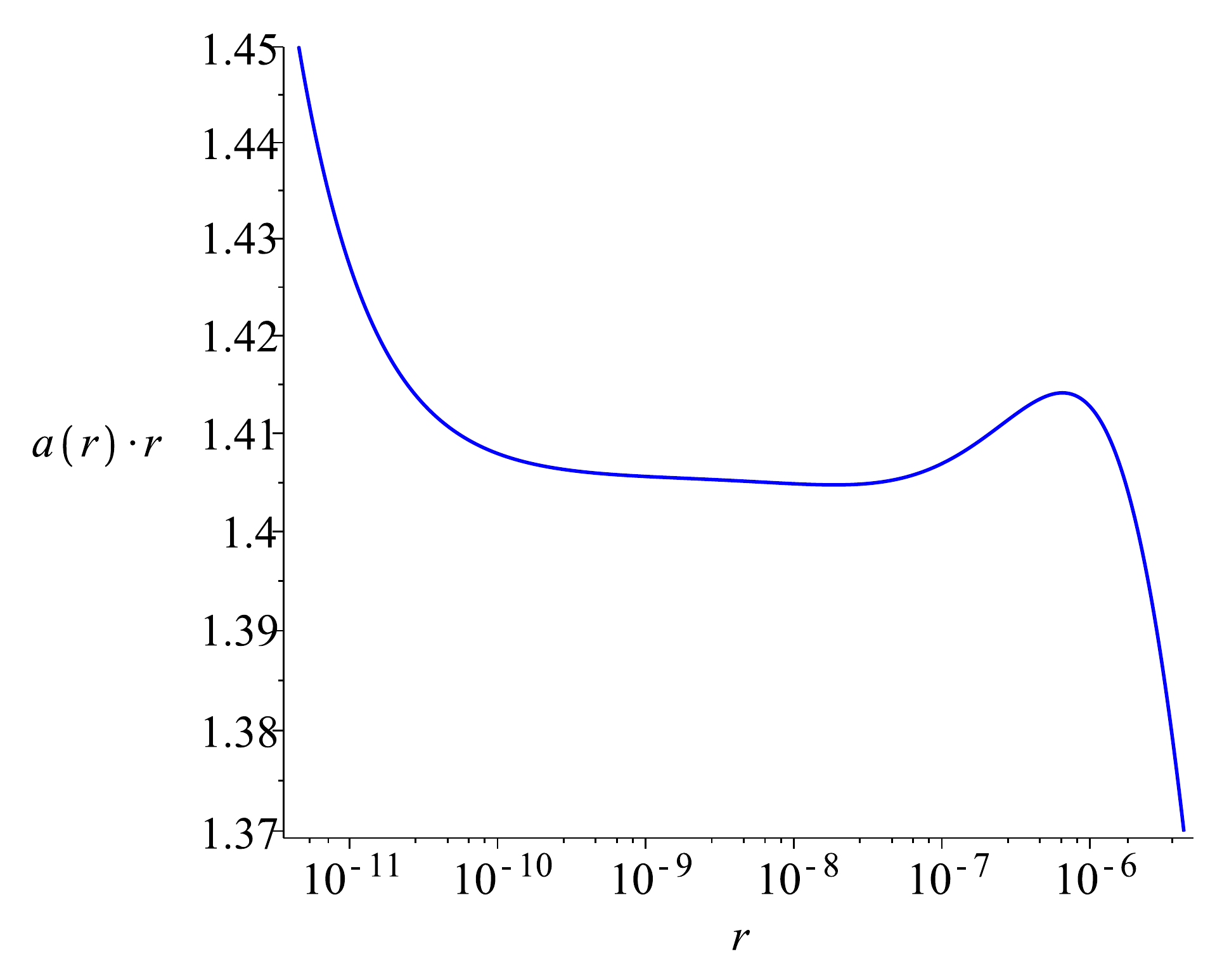}
\caption{\label{fig:a*r}Plot of $a\cdot r$ for solution \#1. The figure on the right is a
magnified view of the Lifshitz region.  The dotted lines
represent the exact AdS$_{2}\times\mathbb{R}^{2}$ solution with $a\cdot r=1$
in the IR and AdS$_{4}$ with $a\cdot r=\sqrt{6}$ in the UV.}
\end{figure}

\begin{figure}[tp]
\centering
\includegraphics[height=6cm]{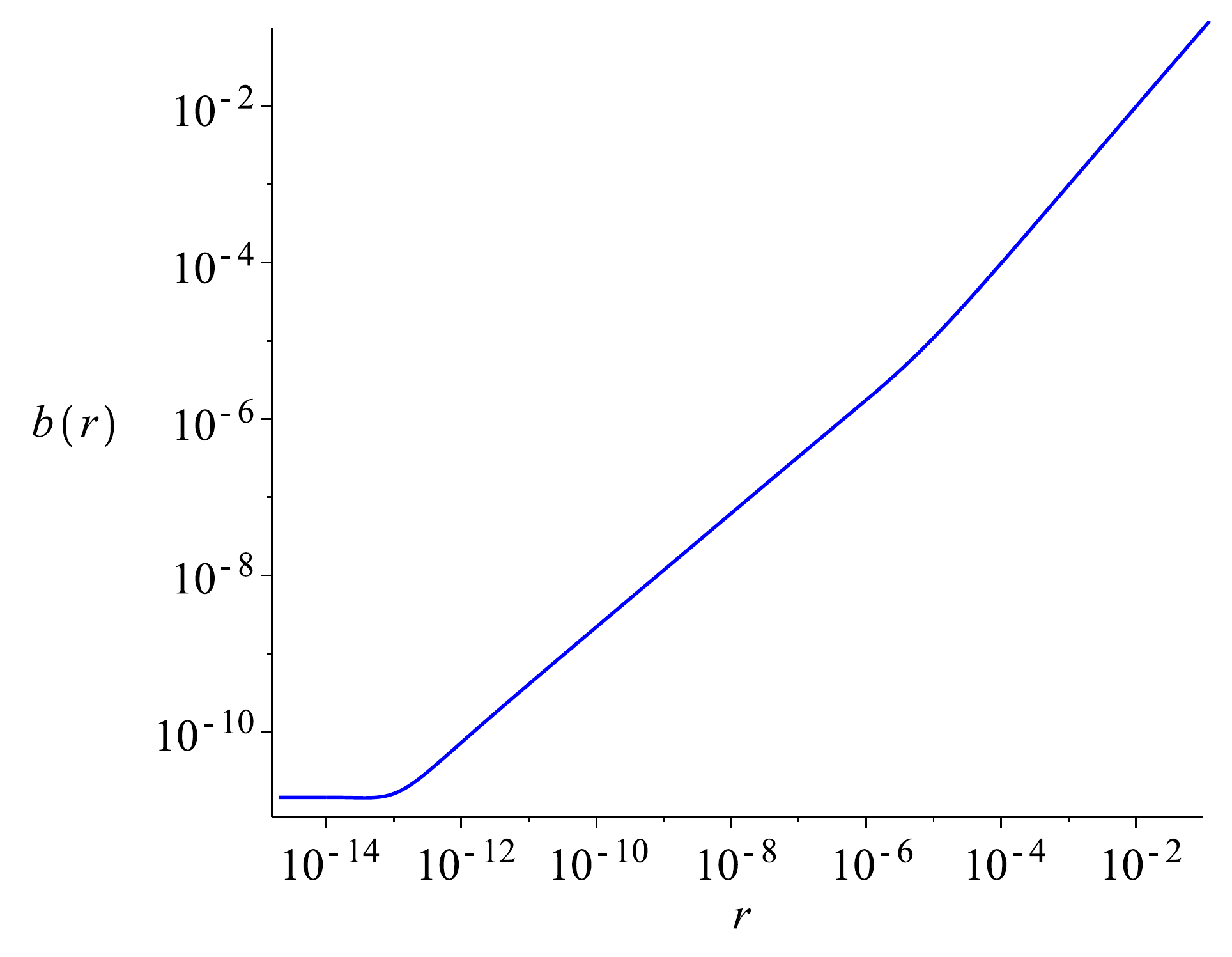}
\hfill
\includegraphics[height=6cm]{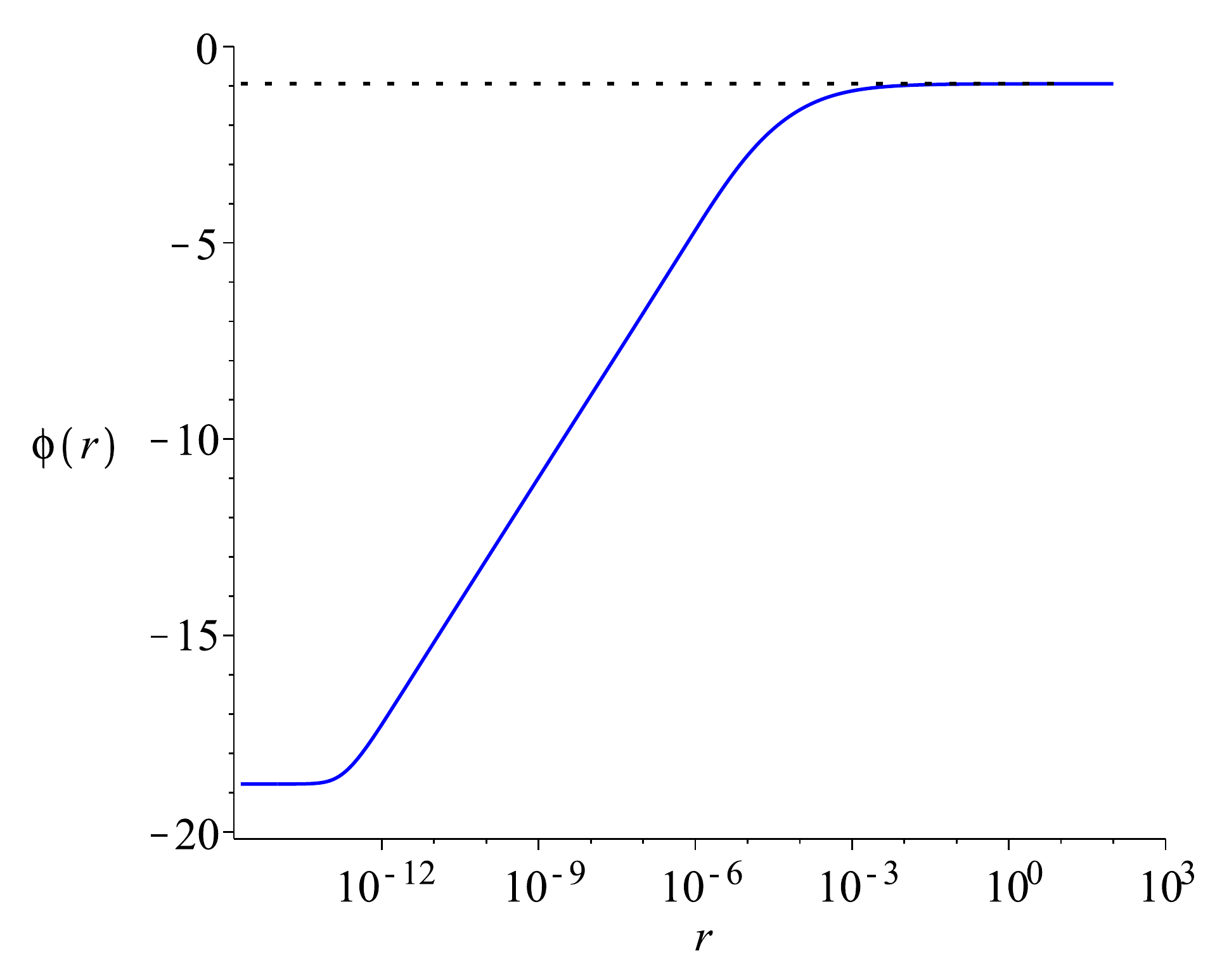}
\caption{\label{fig:b and phi}Plot of the metric function $b$ (left) and the dilaton
$\phi$ (right) for solution \#1.
The dotted line represents the asymptotic value $\phi_{0}$ given
by \eqref{eq:IR value of dilaton}.}
\end{figure}

The parameters and initial conditions of our numerical solutions are
summarized in Table~\ref{tab: initial conditions and parameters}.
Figure~\ref{fig:g00-gxx} shows the evolution of the metric components
$g_{00}$ (black) and $g_{ii}$ (blue) for solution \#1. (The individual
metric functions $a(r)$ and $b(r)$ as well as the dilaton $\phi(r)$ are plotted
in Figures~\ref{fig:a*r} and \ref{fig:b and phi}.)  We chose
to plot ${d\log g_{\mu\nu}}/{d\log r}$ versus $\log r$ so that
power-law relations are clearly visible as horizontal lines. The solution
is asymptotically AdS$_{2}\times\mathbb{R}^{{2}}$ with $g_{00}\propto r^{-2}$,
$g_{ii}\propto r^{0}$ for large $r$. At $r\approx10^{-6}$, the
solution approaches an approximate Lifshitz scaling stage with $g_{00}\propto r^{-2}$,
$g_{ii}\propto r^{2(\tilde{z}-1)}$, where it remains for several
decades. This stage is characterized by an effective scaling parameter
$\tilde{z}_{\mathrm{eff}}\approx0.73$ (or $z\approx3.7$). Notice
that $\tilde{z}_{\mathrm{eff}}$ decreases slowly towards the UV,
as indicated by the slightly positive slope of ${d\log g_{ii}}/{d\log r}$.
This is due to the fact that $e^{\lambda_{2}\phi}$ is small but nonzero:
Effectively, the coupling constant $\alpha$ is reduced, which in
turn increases $\tilde{z}_{\mathrm{eff}}$ (see Figure~\ref{fig:l1(z)}). We expect that as we approach
the attractor, the solution will take the exact form \eqref{eq:Lifshitz solns in new gauge}
with the predicted value of $\tilde{z}\simeq0.71$ for $r\rightarrow0$.
Finally, it is worth mentioning that both $g_{ii}$ and $g_{00}$
initially overshoot slightly before flowing to Lifshitz.

The dilaton starts out at some large negative value $\phi_{\mathrm{UV}}$
for small $r$ and runs towards weak coupling during the scaling stage.
In this intermediate regime, $e^{\lambda_{2}\phi}\ll1$ and $\phi$
grows approximately logarithmically, as in \eqref{eq:dilaton in new gauge}.
As $\phi$ increases, the $e^{\lambda_{2}\phi}$-term becomes more
and more important until at large $r$, the higher derivative corrections
eventually modify the effective potential and stabilize the dilaton
at $\phi_{0}$.

\begin{figure}[tp]
\centering
\includegraphics[height=6cm]{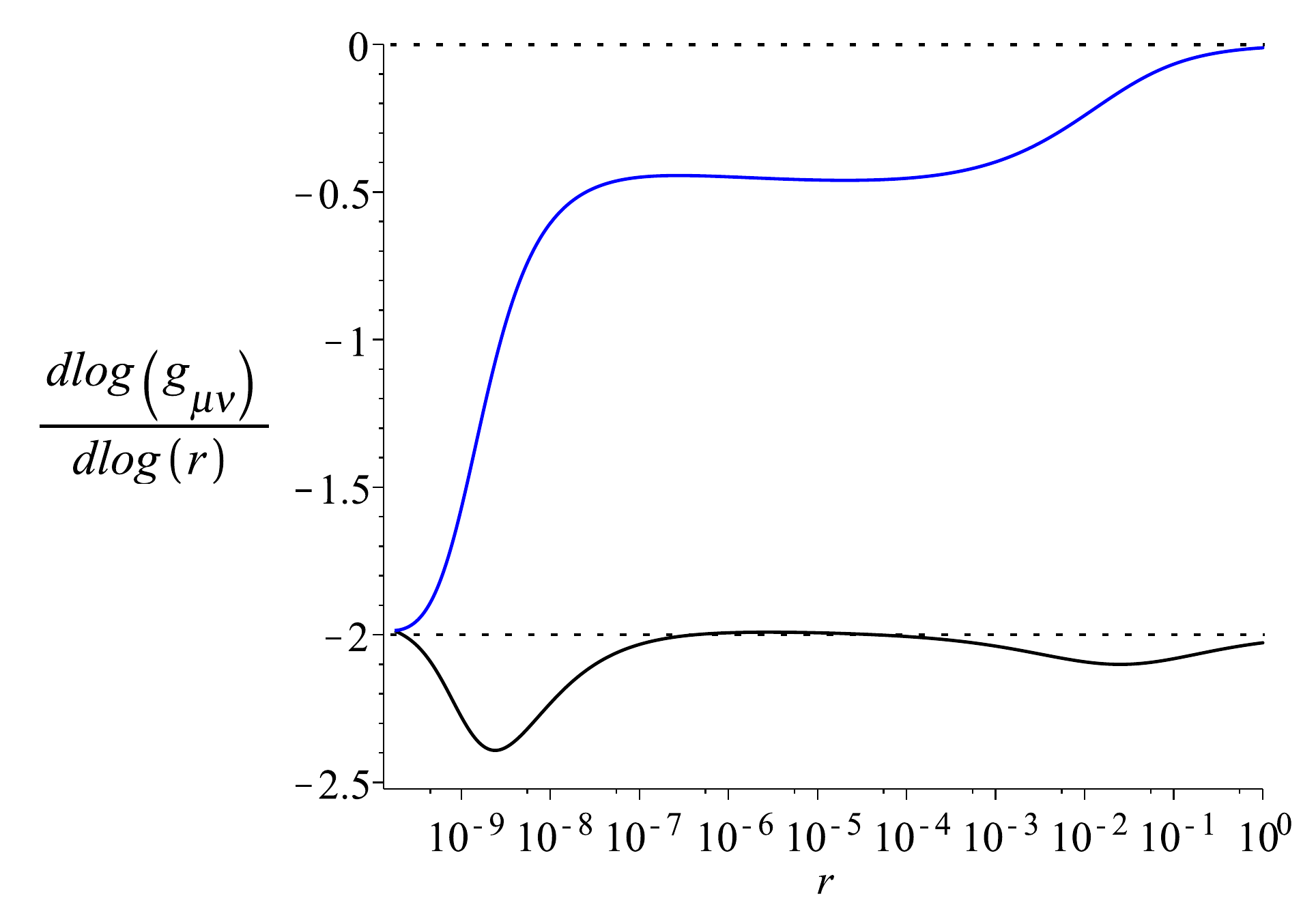}
\caption{\label{fig:alpha>1}Solution \#2 ($\alpha>1$): Plot of the metric
components $g_{00}$ (black) and $g_{ii}$ (blue).}
\end{figure}

In the case of $\alpha\geq1$, there are two possible dynamical exponents
$\tilde{z}_{1}<\tilde{z}_{2}$ (see
Appendix~\ref{sec:Lifshitz solutions in alternative coordinate system}).
There is one additional dilaton perturbation, which we can use to
fix the value of $b^{\prime\prime}$ in the IR. Numerically, we were
only able to find flows from AdS$_{2}\times\mathbb{R}^{2}$ to Lif$_{4}^{\tilde{z}_{2}}$,
with $\tilde{z}_{2}\approx0.78$ ($z_{2}\approx4.4$).  The metric
components for this solution are shown in Figure~\ref{fig:alpha>1}.
The corresponding values for the exact analytical
solution are $\tilde{z}_{1}\approx0.38$ and $\tilde{z}_{2}\approx0.73$.
Although a simple counting of dilaton perturbations would suggest
that there is one irrelevant deformation leading to each of the two
Lifshitz solutions, we were not able to numerically shoot to Lif$_{4}^{\tilde{z}_{1}}$.
It therefore remains unclear whether flows to Lif$_{4}^{\tilde{z}_{1}}$
exist.

\subsection{Flow to $\mathrm{AdS}_{4}$ in the UV}

\begin{figure}[tp]
\centering
\hspace{-0.5cm}\includegraphics[height=5.6cm]{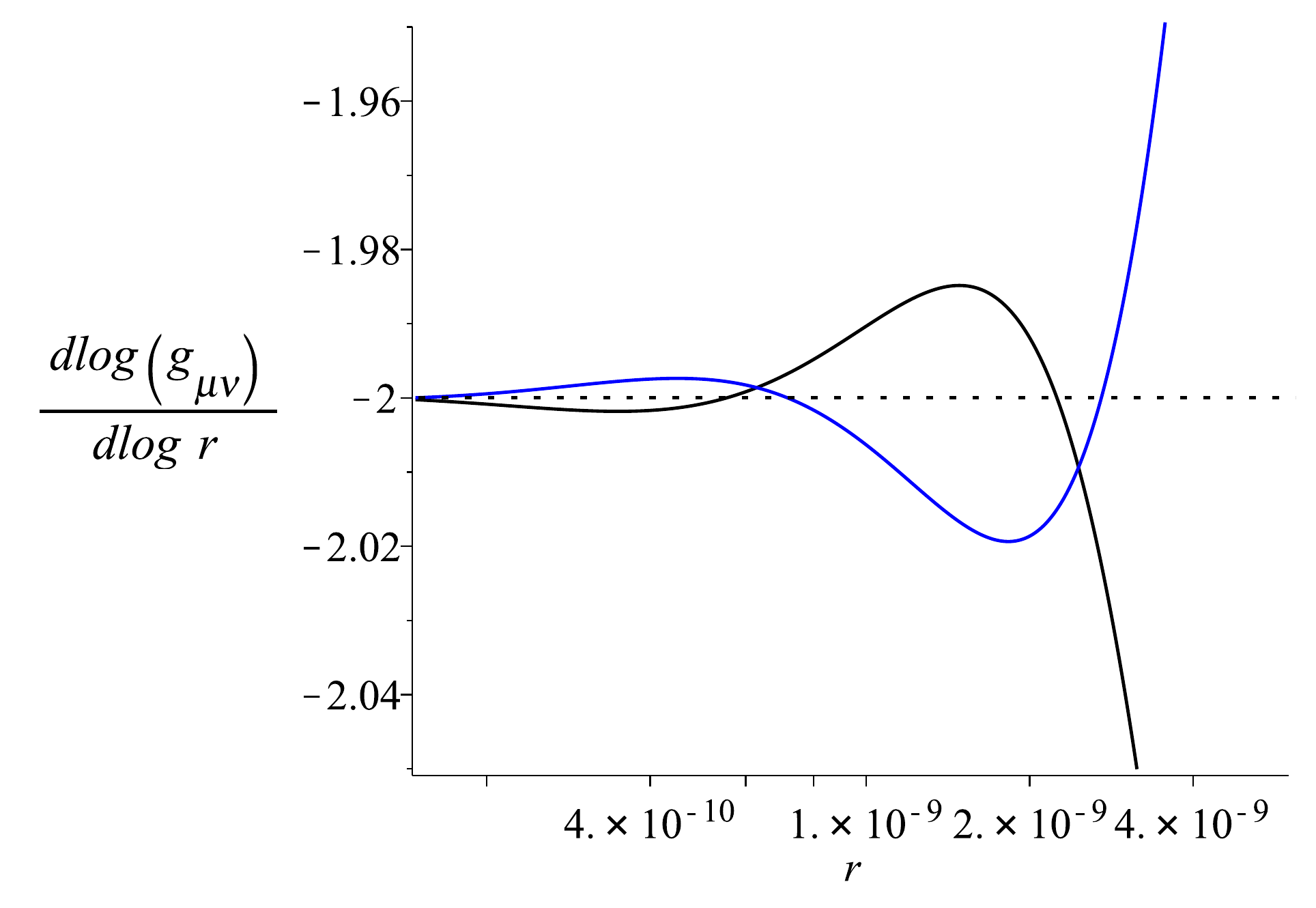}
\hfill
\includegraphics[height=5.6cm]{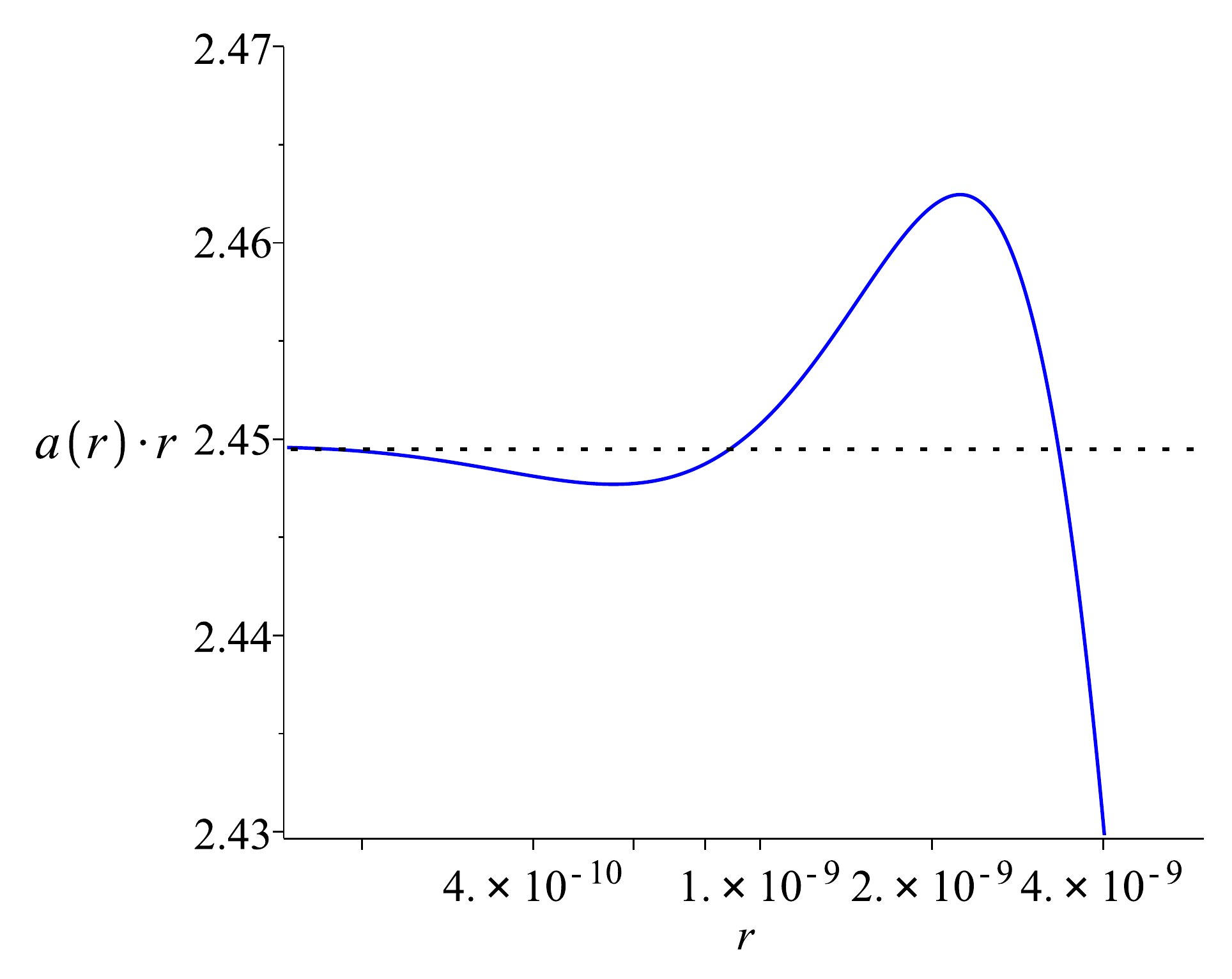}\\
\includegraphics[height=5.6cm]{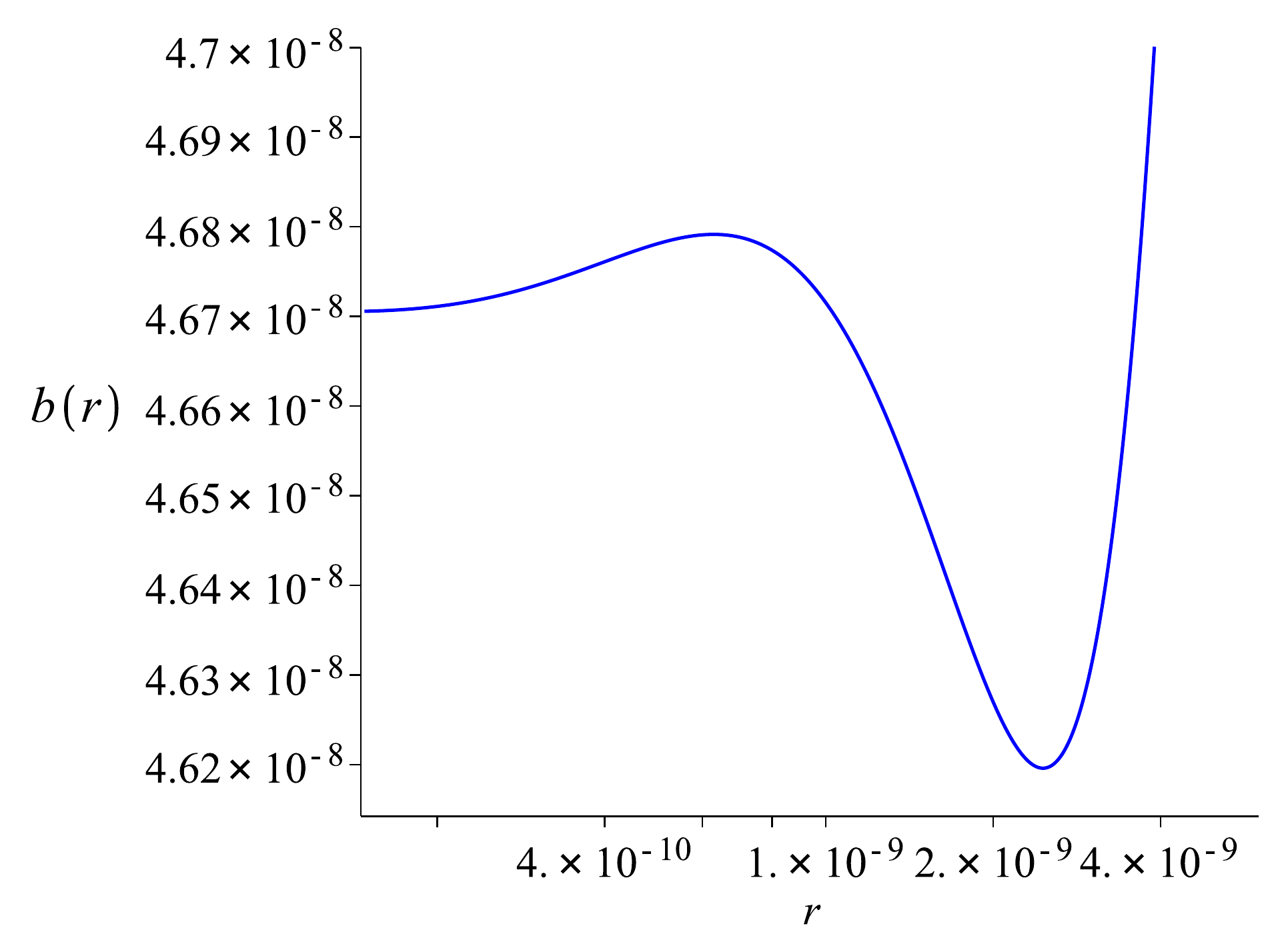}
\hfill
\includegraphics[height=5.6cm]{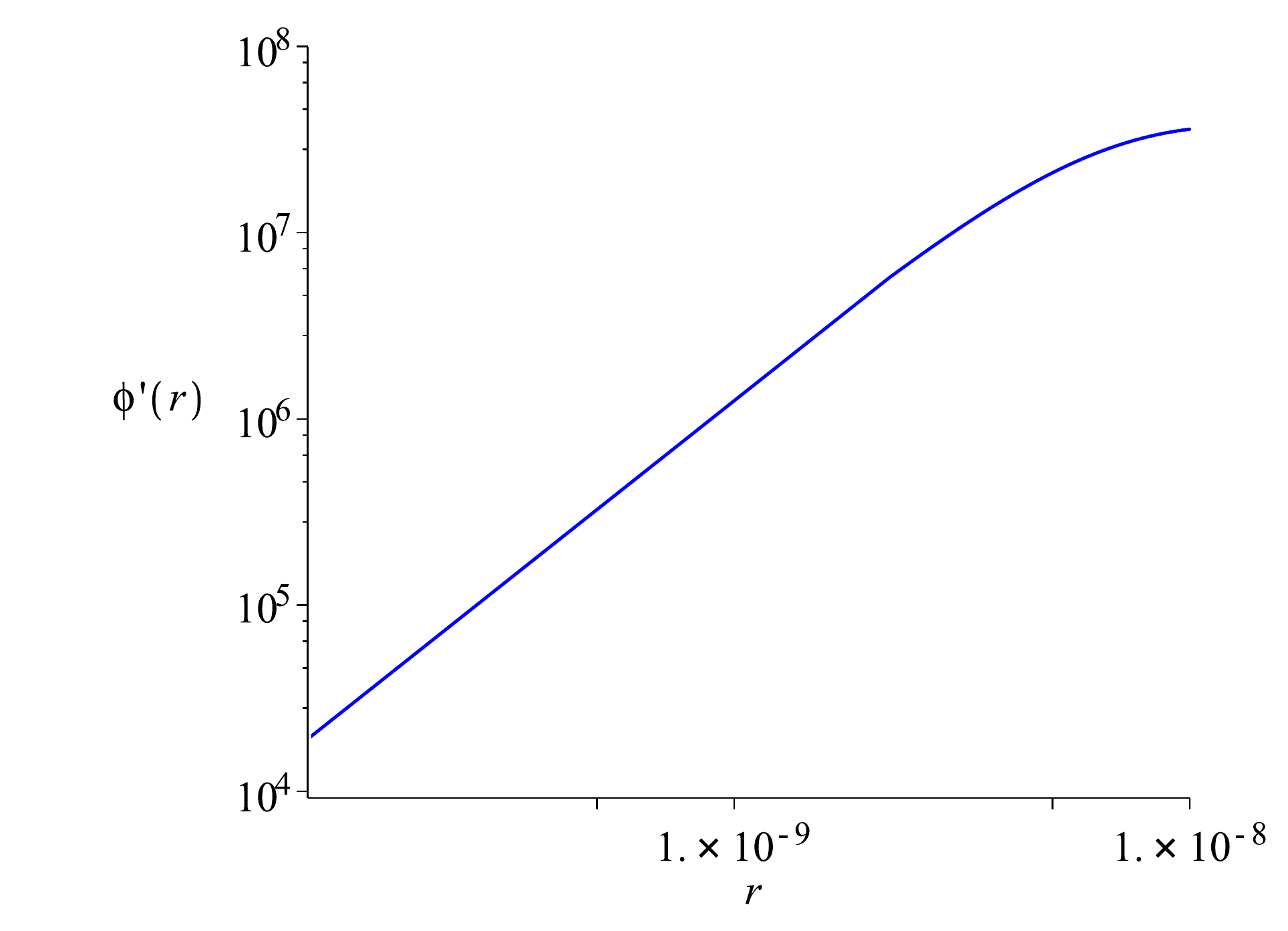}
\caption{\label{fig:UV-asymptotics}Flow to asymptotic AdS$_{4}$ in the deep
UV. The first graph shows the metric components $g_{00}$ (black) and $g_{ii}$ (blue). While the metric functions a and b oscillate according to \eqref{eq:oscillation of a and b},
the dilaton decreases monotonically.}
\end{figure}

Our numerical analysis suggests that the solutions exhibit some interesting
behavior as they approach asymptotic AdS$_{4}$ for $r\rightarrow0$.
It is worthwhile to analyze this asymptotic behavior analytically.
To lowest order, the solution to the linearized equations of motion
is given by
\begin{eqnarray}
a\left(r\right) & = & \frac{\sqrt{6}}{r}\left(1+a_{+}r^{\nu_{+}}+a_{-}r^{\nu_{-}}+a_{3}r^{3}+a_{4}r^{4}+\cdots\right),\nonumber \\
b(r) & = & b_{\mathrm{{\scriptscriptstyle UV}}}+b_{+}r^{\nu_{+}}+b_{-}r^{\nu_{-}}+b_{3}r^{3}+b_{4}r^{4}+\cdots,\nonumber \\
\phi\left(r\right) & = & \phi_{\mathrm{{\scriptscriptstyle UV}}}+\phi_{3}r^{3}+\phi_{4}r^{4}+\cdots,
\end{eqnarray}
where $a_{3}$, $\phi_{3}$, $a_{\pm}$ are free constants and
\begin{eqnarray}
a_{4}&=&\frac{1}{180}\frac{Q^{2}\left(9+2g\left(\phi_{{\scriptscriptstyle \mathrm{UV}}}\right)\right)}{b_{{\scriptscriptstyle \mathrm{UV}}}^{4}f\left(\phi_{{\scriptscriptstyle \mathrm{UV}}}\right)\left(1+2g\left(\phi_{{\scriptscriptstyle \mathrm{UV}}}\right)\right)}, \nonumber \\
b_{3} & = & -2b_{\mathrm{{\scriptscriptstyle UV}}}a_{3},\nonumber \\
b_{4} & = & -\frac{1}{12}\frac{Q^{2}}{\left(1+2g\left(\phi_{\mathrm{{\scriptscriptstyle UV}}}\right)\right)f\left(\phi_{\mathrm{{\scriptscriptstyle UV}}}\right)b_{\mathrm{{\scriptscriptstyle UV}}}^{3}},\nonumber \\
b_{\pm} & = & -\frac{3\left(\nu+1\right)}{2\nu}b_{\mathrm{{\scriptscriptstyle UV}}}a_{\pm},\nonumber \\
\phi_{4} & = & -\frac{1}{12}\frac{Q^{2}f^{\prime}\left(\phi_{\mathrm{{\scriptscriptstyle UV}}}\right)}{b_{\mathrm{{\scriptscriptstyle UV}}}^{4}f\left(\phi_{\mathrm{{\scriptscriptstyle UV}}}\right)^{2}},\nonumber \\
\nu_{\pm} & = & \frac{3}{2}\pm\frac{1}{2}\sqrt{1-\frac{16}{g\left(\phi_{\mathrm{{\scriptscriptstyle UV}}}\right)}}.\label{eq:perturbations around AdS4}
\end{eqnarray}
The leading order perturbations $r^{\nu}$ are purely gravitational.
They survive in the limit of pure Einstein-Weyl gravity (i.e. $Q\rightarrow0$)
\cite{Lu:2012xu}. For $g\left(\phi_{\mathrm{{\scriptscriptstyle UV}}}\right)<16$,
$\nu$ becomes complex and the perturbations oscillate as 
\begin{eqnarray}
a & \sim & r^{\mathrm{Re}\left(\nu\right)-1}\cos\left(\mathrm{Im}\left(\nu\right)\log\left(r\right)+\varphi_{a}\right),\nonumber \\
b & \sim & r^{\mathrm{Re}\left(\nu\right)}\cos\left(\mathrm{Im}\left(\nu\right)\log\left(r\right)+\varphi_{b}\right),\label{eq:oscillation of a and b}
\end{eqnarray}
where $\varphi_{a/b}$ are constant phases. Notice that for $\alpha\rightarrow0$,
the imaginary part of $\nu$ blows up, so these perturbations do not
decouple in the two-derivative limit. Figure~\ref{fig:UV-asymptotics}
shows the asymptotic behavior of one of our numerical solutions (parameter
set \#1). One can clearly see that $a$ and $b$ oscillate according
to \eqref{eq:oscillation of a and b}, while $\phi$ simply decreases
monotonically. 

As it turns out, the oscillating nature of our solutions makes it
necessary to switch to a ``stiff'' method when trying to find exact
numerical solutions in the UV. In addition, the attractor-mechanism
of the Lifshitz stage tends to ``wipe out'' initial conditions,
which makes it more and more difficult to exactly hit AdS$_{4}$ numerically
as the scaling stage gets wider. We therefore content ourselves with
presenting the asymptotic behavior for a solution with a relatively
narrow scaling stage. A more efficient way of studying the UV-asymptotics
would be to directly shoot from the UV.

\section{Discussion}

\label{sec:discussion}

We first showed that Lifshitz
backgrounds are renormalized in the presence of higher derivative
corrections, and in particular Weyl-squared corrections, according to
\eqref{eq:Lifshitz solution, phi}-\eqref{eq:Lifshitz solution, Lambda}. However, the exceptions to
this are solutions of conformal gravity,
which may be obtained in the formal limit $\alpha_{\mathrm{{\scriptscriptstyle W}}}\rightarrow\infty$.
The variation of $C_{\mu \nu \rho \sigma }^2$ is proportional to the Bach-tensor 
\begin{equation}
B_{\mu\nu}=\left(\nabla^{\rho\sigma}+\frac{1}{2}R^{\rho\sigma}\right)C_{\mu\rho\nu\sigma}.
\end{equation}
Notice that the Bach-tensor vanishes identically for Einstein metrics.
Moreover, since $B_{\mu\nu}$ is a conformal tensor, it also vanishes on
spacetimes that are conformally Einstein.
Hence, these backgrounds are not renormalized. For the solutions
of the form \eqref{eq:lifbulk}, the cases $z=0$, $z=1$ (AdS) and, in four dimensions,
$z=4$ are conformally related to Einstein metrics%
\footnote{They are, in fact, conformally Ricci flat.%
} via $g_{\mu\nu}\mapsto e^{-2({2+z^{2}})/({2+z})}g_{\mu\nu}$ (for
$d=3$) and are therefore protected against renormalization.

We then demonstrated in a toy model that higher curvature corrections, such as those that
arise from the string $\alpha'$ expansion, may resolve the Lifshitz horizon into
AdS$_2\times \mathbb R^2$.  In particular, we have constructed numerical flows from AdS$_4$
to an intermediate Lifshitz region and finally to AdS$_2\times\mathbb R^2$ in the deep IR
in the Einstein-Maxwell-dilaton system with a four-derivative correction of the form
\begin{equation}
\delta\mathcal L=\fft34(\alpha+\beta e^{\lambda_2\phi})C_{\mu\nu\rho\sigma}^2.
\end{equation}
The dilaton coupling $\beta$ is introduced to stabilize the dilaton, so that an emergent
AdS$_2\times\mathbb R^2$ may appear in the IR.

The existence of flows to AdS$_2\times\mathbb R^2$ is not universal, but depends on
the parameters $\alpha$, $\beta$, $\lambda_1$ and $\lambda_2$.  For $\alpha<1$, there
is at most one irrelevant dilaton perturbation that can induce a flow from
AdS$_2\times\mathbb R^2$ in the deep IR to an intermediate Lifshitz region.  We have
presented a numerical example of such a flow for $\alpha=0.9$.  On the other hand,
for $\alpha\ge1$, if any irrelevant dilaton perturbations exist, then they necessarily come as
a pair.  Furthermore, in this case there are two possible dynamical exponents, $\tilde z_1$
and $\tilde z_2$ (where we take $\tilde z_1<\tilde z_2$), allowed in the Lifshitz region.
We have constructed a numerical example for $\alpha=3$ that flows from 
$\mathrm{AdS}_{2}\times\mathbb{R}^{2}$ in the deep IR to an intermediate
$\mathrm{Lif}_{4}^{\tilde{z}_{2}}$.  However, we were unable to find numerical flows to
$\mathrm{Lif}_{4}^{\tilde{z}_{1}}$.  It remains unclear whether such flows are possible. From a
simple counting of irrelevant perturbations, we expect that these flows should
indeed exist. In any case, the natural question that arises is whether or not the additional
irrelevant perturbation that appears for $\alpha\geq1$ leads to an interesting geometry. To
make a definitive statement about the flows that are allowed, a study of perturbations around
the different Lifshitz backgrounds would be required. A similar analysis was carried out
for the massive vector case in \cite{Braviner:2011kz,Liu:2012wf}. It would also be interesting
to see if one can find numerical solutions that interpolate between the two Lifshitz solutions.

It is also with noting that, for a certain choice of parameters, we found irrelevant perturbations
that oscillate around AdS$_{2}\times\mathbb{R}^{2}$. We have discarded such cases, as
our intuition from two-derivative theories suggests that this should be taken as
a sign of a dynamical instability. However, we would have to perform a more detailed
analysis to see whether the existence of such oscillatory perturbations actually leads to
an instability.  If in these cases the IR geometry is truly unstable, this raises the question of
what the geometry would decay into and consequently what the true ground
state of the theory is. Another source of instability that we did not consider here is the
formation of striped phases \cite{Donos:2011qt,Donos:2011pn,Cremonini:2012ir}.

Of course, it would be desirable to explore whether a realistic string model would
lead to either $\alpha'$ or string loop corrections of the form needed to resolve the Lifshitz
horizon.  The $\alpha'$ corrections extend beyond the gravitational sector, and for
example may include $RF^2$ terms at the four derivative level.  Even in the gravitational
sector, one would expect to have a more general form of the four-derivative corrections,
similar to (\ref{eq:hd-Lagrangian, new basis}), but also with possible dilaton couplings.
We expect that the mechanism to resolve the Lifshitz singularity in
the IR will also work in the more general case with $\alpha_{\mathrm{{\scriptscriptstyle R}}}\ne0$
and $\alpha_{\mathrm{{\scriptscriptstyle GB}}}\neq0$.
However, the smooth flow to AdS$_{4}$ in the UV observed here relies on the fact that the
Weyl tensor vanishes quickly enough so that it does not source the
dilaton for small $r$. It is unclear whether the UV asymptotics would
remain unchanged for \textit{generic} higher derivative corrections.

In the case of an electrically charged brane considered here, there
is another source of corrections that might modify the UV dynamics:
Since the dilaton runs towards strong coupling, we expect
quantum corrections to the gauge kinetic function $f\left(\phi\right)$
to become important and modify the effective potential in this regime.
In addition, there is a priori no reason why magnetic solutions should not be equally
sensitive to $\alpha^{\prime}$-corrections. We therefore expect our
mechanism to be relevant also in the magnetic case. Since in this
case the dilaton runs towards strong coupling in the IR, a consistent approach would be 
to consider both $\alpha^{\prime}$ and quantum corrections at the same time.

We expect that our analysis can be easily extended to geometries
with hyperscaling violation. These backgrounds can be parametrized
by a metric of the form 
\begin{equation}
ds_{d+1}^{2}=e^{2\gamma r}(-e^{2zr/L}dt^{2}+e^{2r/L}d\vec{x}^{2}+dr^{2}).
\end{equation}
For $\gamma\neq0$ this metric is invariant under the scale transformation
\eqref{eq:Lifshitz scaling symmetry} and \eqref{eq:Lss2} only up to a
rescaling of $ds$. One may construct solutions of this type by choosing an
exponential potential for the dilaton, which as a result runs linearly with $r$.
Flows to AdS$_2\times\mathbb R^2$ were constructed using a quantum
corrected gauge kinetic function $f(\phi)$ in \cite{Bhattacharya:2012zu}.

Finally, although an emergent $AdS_{2}\times\mathbb{R}^{2}$ geometry
provides a non-singular resolution of the Lifshitz scaling solution, the presence
of a non-contracting transverse $\mathbb R^2$ leads to non-zero entropy at
zero temperature in the dual non-relativistic system.  A more realistic situation
where the entropy vanishes at zero temperature may potentially be obtained
by flowing into AdS$_4$ in the deep IR.  Thus one may imagine constructing
flows from AdS$_4$ to Lifshitz to AdS$_4$.  This would be a special case of
an AdS to AdS domain wall solution, in which case the holographic $c$-theorem
would apply.  It would be interesting to see whether such flows may be
constructed in a toy model admitting AdS$_4$ solutions with two distinct AdS radii.


\section*{Acknowledgments}

We wish to thank S. Cremonini, H. Elvang, S. Roland and P. Szepietowski for useful
discussions. This work was supported in part by the US Department of Energy under
grant DE-SC0007859.


\appendix

\section{Metric ansatz and curvature}
\label{app:A}

Here we provide the curvature components used in the derivation of the Lifshitz solution in
section~\ref{sec:hdlif}. Although not needed for the Lifshitz case, we consider slightly more
general metrics of the form
\begin{equation}
ds^{2}=-e^{2b_{0}(r)}dt^{2}+dr^{2}+\sum_{i=1}^{d-1}e^{2b_{i}(r)}\left(dx^{i}\right)^{2}.
\label{eq:scaling metric ansatz}
\end{equation}
The nonvanishing curvature terms are
\begin{eqnarray}
R_{\phantom{\rho}\sigma\mu\nu}^{\rho} & = & \delta_{\nu}^{\rho}\eta_{\mu\sigma}b_{\nu}^{\prime}b_{\sigma}^{\prime}e^{2b_{\sigma}}-(\mu\leftrightarrow\nu),\label{eq:Riemann}\\
R_{\phantom{\rho}\mu r\nu}^{r}&=&-\eta_{\mu\nu}e^{2b_{\nu}}\big(b_{\nu}^{\prime\prime}+\left(b_{\nu}^{\prime}\right)^{2}\big)\\
R_{rr} & = & -\sum_{\lambda}(b_{\lambda}^{\prime\prime}+(b_{\lambda}^{\prime})^{2}),\label{eq:R_rr}\label{eq:Riemann2r}\\
R_{\mu\nu} & = & -\eta_{\mu\nu}e^{2b_{\nu}}\bigg(b_{\nu}^{\prime\prime}+b_{\nu}^{\prime}\sum_{\lambda}b_{\lambda}^{\prime}\bigg),\label{eq:R_munu}\\
R & = & -\sum_{\lambda}\big(2b_{\lambda}^{\prime\prime}+(b_{\lambda}^{\prime})^{2}\big)-\bigg(\sum_{\lambda}b_{\lambda}^{\prime}\bigg)^{2},\label{eq:R}
\end{eqnarray}
where $\mu,\nu=0,\ldots,d-1$, and repeated indices are not summed over
unless explicitly stated. The Lifshitz solution is given by
\begin{eqnarray}
b_{0}\left(r\right) & = & zr\;,\qquad b_{i}(r)=r\;,\qquad z>1,\label{eq:b's for Lifshitz solutions}
\end{eqnarray}
and the case $z=1$ corresponds to $AdS_{d\text{+1}}$. For this
class of solutions, we have
\begin{eqnarray}
R_{\phantom{0}0r0}^{r} & = & z^{2}e^{2zr},\nonumber \\
R_{\phantom{r}irj}^{r} & = & -\delta_{ij}e^{2r},\nonumber \\
R_{\phantom{0}i0j}^{0} & = & -\delta_{ij}ze^{2r},\nonumber \\
R_{\phantom{i}jkl}^{i} & = & (\delta_{l}^{i}\delta_{jk}-\delta_{k}^{i}\delta_{jl})e^{2r},\nonumber \\
R_{\text{00}} & = & z(z+d-1)e^{2zr},\nonumber \\
R_{rr} & = & -(z^{2}+d-1),\nonumber \\
R_{ij} & = & -\delta_{ij}(z+d-1)e^{2r},\nonumber \\
R & = & -(z^{2}+d-1+(z+d-1)^{2}).
\end{eqnarray}

\section{Lifshitz solutions in alternative gauge}

\label{sec:Lifshitz solutions in alternative coordinate system}
\begin{figure}[tp]
\centering
\includegraphics[height=6cm]{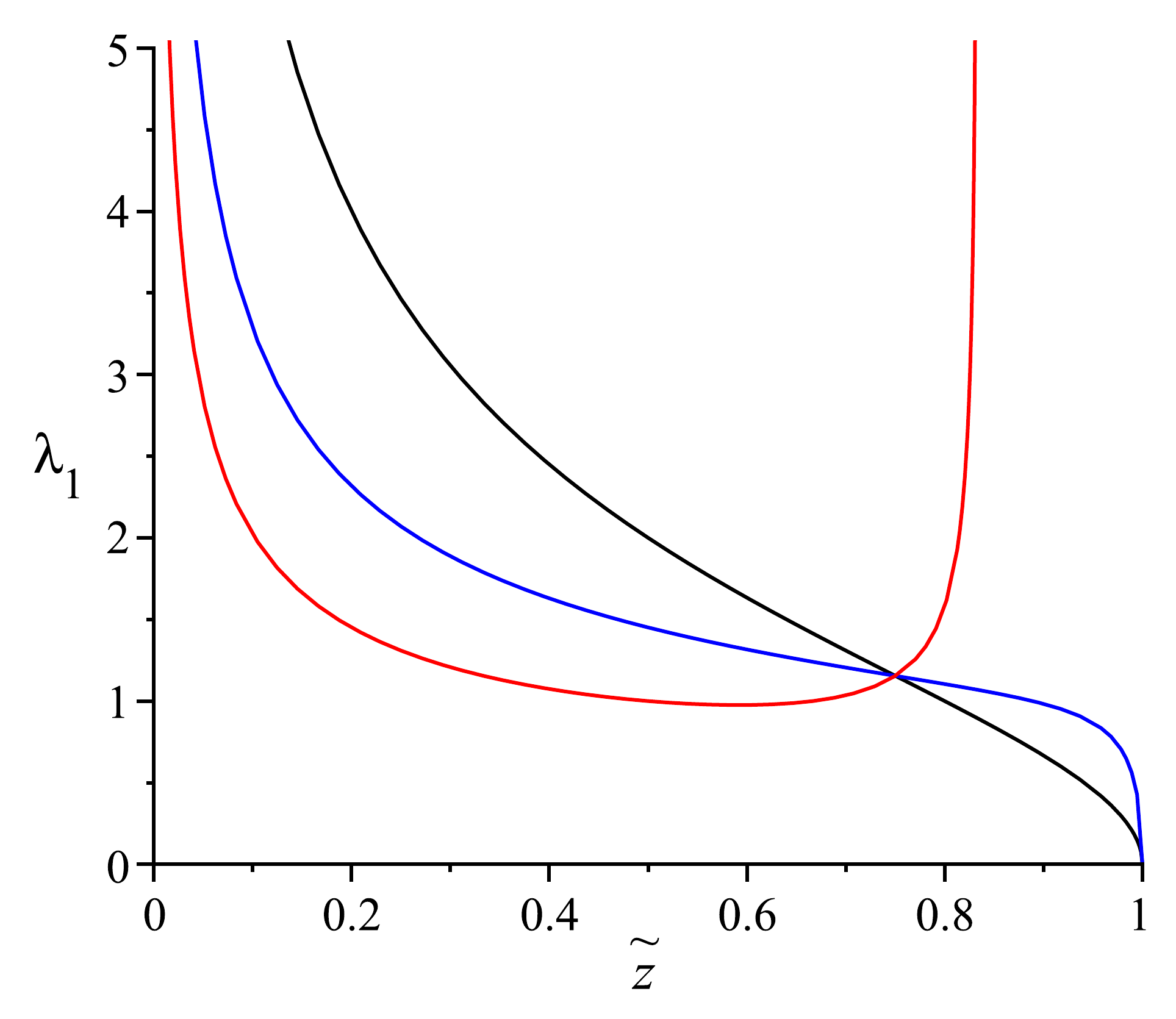}
\caption{\label{fig:l1(z)}Plot of $\lambda_{1}\left(\tilde{z}\right)$ for
$\alpha=0$ (black), $\alpha=0.9$ (blue) and $\alpha=3$ (red). }
\end{figure}

In
our numerical analysis, we chose the parametrization \eqref{eq:metric parametrization}
for the metric, which is different from \eqref{eq:scaling metric ansatz}.
In this gauge, the Lifshitz metric of section \ref{sec:hdlif} takes
the form:
\begin{eqnarray}
ds^{2} & = & \frac{1}{r^{2}}\left(-dt^{2}+dr^{2}+r^{2\tilde{z}}\left(dx^{2}+dy^{2}\right)\right).
\end{eqnarray}
The scaling parameters are related via $z=\left(1-\tilde{z}\right)^{-1}$.
Furthermore,
\begin{eqnarray}
\phi & = & \frac{4\left(1-\tilde{z}\right)}{\lambda_{1}}\log r+C,\label{eq:dilaton in new gauge}\\
Q^{2}e^{-\lambda_{1}C} & = & \left(\frac{3}{2}-\tilde{z}\right)\tilde{z}\left(1-4\alpha\left(\tilde{z}-\frac{3}{4}\right)\right),\\
\Lambda & = & 2\left(\frac{3}{2}-\tilde{z}\right)\left(2-\tilde{z}\right)+4\alpha\tilde{z}\left(1-\tilde{z}\right)\left(\frac{3}{4}-\tilde{z}\right),\\
\lambda_{1}^{2} & = & \frac{4\left(\frac{3}{2}-\tilde{z}\right)\left(1-\tilde{z}\right)}{Q^{2}e^{-\lambda_{1}C}}=\frac{1-\tilde{z}}{\tilde{z}\left(\frac{1}{4}-\alpha\left(\tilde{z}-\frac{3}{4}\right)\right)}.\label{eq:l1^2(ztilde)}
\end{eqnarray}
It is straightforward to show that $\lambda_{1}^{2}\left(\tilde{z}\right)$
has a local minimum at 
\begin{equation}
\tilde{z}_{\pm}=1\pm\frac{1}{2}\sqrt{1-\frac{1}{\alpha}}\;,
\end{equation}
provided that $\alpha\geq1$. In this case there are two different
scaling parameters $\tilde{z}_{1}<\tilde{z}_{2}$ for any given $\lambda_{1}^{2}$
(away from the minimum) (see Figure \ref{fig:l1(z)}). Notice also
that $\lambda_{1}^{2}$ blows up for $\tilde{z}_{\star}={3}/{4}+{1}/({4\alpha})$,
which is within the range of physical solutions for $\alpha\geq1$
only. To summarize, the possible ranges for the parameters are:
\begin{eqnarray}
\alpha<1:\quad & 0\leq\lambda_{1}<\infty\;, & \quad0<\tilde{z}\leq1,\nn\\
\alpha\geq1:\quad & \lambda_{\mathrm{min}}\leq\lambda_{1}<\infty\;, & \quad0<\tilde{z}<\tilde{z}_{\star},
\end{eqnarray}
where $\lambda_{\mathrm{min}}\equiv\lambda_{1}\left(\tilde{z}_{-}\right)$.

\section{Irrelevant Perturbations}

\label{sec:Irrelevant Perturbations}There are two ways in which the
exponent of the dilaton perturbations, $\nu$ may become complex: 
\begin{enumerate}
\item The smaller square-root in \eqref{eq:Ctilde solution} becomes imaginary.
This happens when 
\begin{equation}
\lambda_{2}=\frac{1}{\alpha\lambda_{1}}\left(-\left(\lambda_{1}-\frac{2}{\sqrt{3}}\right)^{2}+\frac{2}{3}\right).\label{eq:lambda2 curve1}
\end{equation}

\item Even if the small root is real-valued, $\widetilde{\nu}^{2}$ may
still cross zero, which happens at
\begin{eqnarray}
\kern-3em\lambda_{2} & = & \frac{1}{\lambda_{1}\left(11\alpha^{2}-19\alpha+8\right)}\bigg[\frac{4}{3}\left(1-\alpha\right)\lambda_{1}^{2}+\frac{11}{8}\alpha-1\nonumber \\
 &  &\kern9em \pm\frac{3}{2}\left(\left(\alpha-1\right)^{2}\lambda_{1}^{4}-\frac{1}{2}\left(11\alpha^{2}-19\alpha+8\right)\lambda_{1}^{2}+\left(\frac{11}{12}\alpha-\frac{2}{3}\right)^{2}\right)^{\frac{1}{2}}\bigg].\nn\\
 \label{eq:lambda2 curve2}
\end{eqnarray}

\end{enumerate}

To find out when the dilaton perturbations are irrelevant, i.e.
$\widetilde{\nu}^{2}={1}/{4}$, notice that $\widetilde{\nu}^{2}-{1}/{4}$
can only change its sign as we go from case 1) to case 2) in
\eqref{eq:two different cases for alpha (<>1)}.
As a consequence, irrelevant perturbations will stay irrelevant as
long as $\alpha{\lambda_{2}}/{\lambda_{1}}\gtrless1$. In practice,
it is therefore easiest to plot the curves \eqref{eq:lambda2 curve1}/\eqref{eq:lambda2 curve2}
and determine the number of irrelevant perturbations numerically,
making use of continuity arguments (see Figures~\ref{fig:contours1}
and \ref{fig:contours2}).


\end{document}